\documentclass[fleqn,usenatbib,useAMS]{mnras}

\usepackage{newtxtext,newtxmath}

\usepackage[T1]{fontenc}

\DeclareRobustCommand{\VAN}[3]{#2}
\let\VANthebibliography\thebibliography
\def\thebibliography{\DeclareRobustCommand{\VAN}[3]{##3}\VANthebibliography}

\usepackage{graphicx}	
\usepackage{amsmath}	
\usepackage{comment}
\usepackage{hyperref}
\usepackage{cleveref}
\usepackage{subfig}
\usepackage{url}
\usepackage{soul}
\newcommand{\df}[2]{ \frac{\partial {#1}}{\partial {#2}} }
\renewcommand{\l}{\left}
\renewcommand{\r}{\right}

\title[Probing twin stars with $f$-modes]{Probing hadron-quark phase transition in twin stars using $f$-modes}

\author[B. K. Pradhan et al.]{
Bikram K. Pradhan,$^{1}$\thanks{E-mail: bikramp@iucaa.in (KTS)}
Debarati Chatterjee,$^{1}$
David Edwin Alvarez-Castillo $^{2,3,4}$
\\
$^{1}$Inter-University Centre for Astronomy and Astrophysics,  Pune, India\\
$^{2}$The Henryk Niewodnicza\'nski Institute of Nuclear Physics,
Polish Academy of Sciences, ul. Radzikowskiego 152, PL-31-342 Krak\'ow, Poland
\\
$^{3}$Incubator of Scientific Excellence - Centre for Simulations of Superdense Fluids, University of Wroclaw, plac Maksa Borna 9, PL-50204 Wroclaw, Poland
\\
$^{4}$Facultad de Ciencias Físico Matemáticas, Universidad Autónoma de Nuevo León, Av. Universidad S/N, C.U., 66455 San Nicolás de los Garza, N.L., Mexico
}

\date{Accepted XXX. Received YYY; in original form ZZZ}

\pubyear{2015}

\begin{document}
\label{firstpage}
\pagerange{\pageref{firstpage}--\pageref{lastpage}}
\maketitle

\begin{abstract}
Although it is conjectured that a phase transition from hadronic to deconfined quark matter in the ultrahigh-density environment of Neutron Stars (NS), the nature of phase transition remains an unresolved mystery.{Furthermore, recent efforts reveal that the finite surface tension effects can lead to a mixed phase with different geometric shapes (so-called "pasta" phases), leading to a smooth phase transition from hadronic to quark matter in the NS interior. Depending on whether there is a strong or a pasta-induced smooth first-order phase transition, one may expect a third family of stable, compact stars or "twin stars" to appear}, with the same mass but different radii compared to NSs. The possibility of identifying twin stars using astrophysical observations has been a subject of interest. This study investigates the potential of probing the nature of the hadron-quark phase transition through future gravitational wave (GW) detections from fundamental ($f$-) mode oscillations in Neutron Stars. Using a newly developed model that parametrizes the hadron-quark phase transition with ``pasta phases," we calculate $f$-mode characteristics within a full general relativistic framework. We then use Universal Relations in GW asteroseismology to derive stellar properties from the detected mode parameters. Our findings suggest that detecting GWs from $f$-modes with third-generation GW detectors offers a promising scenario for the existence of twin stars. However, we also estimate various uncertainties in determining the mode parameters and conclude that these uncertainties make it more challenging to identify the nature of the hadron-quark phase transition.

\end{abstract}




\section{Introduction}
\label{sec:intro}

Since their discovery, the interior composition of neutron stars (NS) remains one of the most intriguing unanswered questions in astrophysics. Although terrestrial experiments provide hints about the behaviour of matter at high densities, the densities in the core of neutron stars surpass those of laboratories by several orders of magnitude.
This makes compact objects the ideal environment to probe matter under extreme conditions of low temperature and high densities not accessible to terrestrial experiments. Quantum chromodynamics (QCD) predicts the appearance of strange matter at high densities via a first-order phase transition from hadronic matter. It is therefore conjectured that deconfined quark matter may appear as a stable component in the high-density environment of the neutron star inner core~\citep{Christian2023,Muses2023,Takatsy2023,Shirke_2023,Blaschke2020}.  The phase diagram of QCD corresponding to dense nuclear matter in the hot and low density regime has been explored by laboratory experiments, namely relativistic nuclear collisions in the Large Hadron Collider (LHC) as well as by numerical solutions of Lattice QCD. The result is a smooth phase transition or crossover. Both the intermediate and lower temperature and density regions  are not very well studied experimentally or theoretically, therefore the nature of the phase transition is quite uncertain~\citep{Muses2023}. 
{In particular, compact star matter located
at the low-temperature and high-density region of the QCD phase diagram can potentially feature a phase transition from hadronic matter to deconfined quark matter. This phase transition is conjectured to be a first order which might or might not feature geometrical shapes at the interface. Alternatively, there are proposals that state that the hadron-quark phase transition ought to be a crossover, see for instance~\cite{Baym:2017whm,Sotani:2023zkk}. Within this work, the first possibility namely the QCD first order phase transition is addressed.}
\\
 

Although equations of QCD remain unsolved at energy scales relevant for describing NS matter, it is possible to model the NS interior via equations of state (EOS) based on microscopic or phenomenological density functional theories. Through the EOS, the NS interior composition can then be connected to global NS properties, such as its mass or radius in electromagnetic observations or tidal deformation of the components of a binary system during a merger from gravitational wave emission. It is expected that the appearance of strange degrees of freedom would result in a ${\it softer}$ EOS and correspondingly to a lower maximum NS mass. This can then be compared to mass measurements from NSs in binary systems to test whether a particular EOS can support the observed maximally massive NS. 
Moreover, neutrinos from core-collapse supernova, from accreting compact stars, and  gravitational wave signals from their binary mergers produced at the postmerger phase can potentially identify and constrain the properties of a first order phase transition of hadronic matter into quark matter~\citep{Largani:2023kjx,Largani:2023oyk,Bauswein:2018bma}.
\\

Further, depending on the nature of the hadron-quark phase transition, distinct branches in the NS mass-radius relation may appear~\citep{Glendenning1992,Blaschke2020,Zacchi2017,Espino2022}. In recent years, a significant amount of investigations have been carried out connecting the astrophysical observations and twin stars~\citep{Christian2022,Christian_2020,Tsaloukidis2023,Gloria2019}.  The detection of compact star mass twins, i.e., two stars of about the same mass but different radii, would be a smoking gun evidence for a strong first order phase transition. On the contrary, a smooth phase transition that might be caused by the appearance of non-homogeneous matter or ${\it pasta}$ phases at the quark-hadron interface might not result in a third branch of compact stars in the mass-radius diagram. Studies carried out in~\citep{Ayriyan,Blaschke:2018pva,Blaschke20206,Maslov2019} reveal an effective assessment of the amount of quark-hadron pasta that could support the neutron star while preserving the third, disconnected compact star branch.
\\

Recently there has been significant interest in trying to identify whether hadron-quark transition occurs via either a {${\it rapid}$ or ${\it slow}$ process~\citep{Pereira2018} at the boundary between pure phases (accompanied by an energy density jump sometimes resulting in compact stars twins~\citep{Landry2022,David2021}). In this work we also consider a phase transition via a mixed phase~\citep{Abgaryan2018,Ayriyan,AyriyanEPJ2018,Ayriyan2021b}.} Several phenomenological interpolation schemes were proposed to mimic the hadron-quark mixed phase in compact stars via geometrical structure of different shapes denoted as ``pasta phases"~\citep{Abgaryan2018,Ayriyan2021b}, and their effect on properties of hybrid stars were investigated.  \citep{Pereira2018} studied whether the nature of the phase transition can be assessed from future detections of global NS parameters such as mass, radius and tidal deformability. Landry and Chakraborty~\citep{Landry2022} explored the possibility of distinguishing neutron stars from compact twins from future detections of tidal deformability of binary NS mergers.
\\

The LIGO-VIRGO collaboration has, until now, directly witnessed gravitational waves (GW) from binary neutron stars (BNS), and the discovery of the BNS event GW170817 along with the electromagnetic counterpart opens an exciting new chapter in multi messenger astronomy~\citep{AbbottPRL121,AbbottPRL119,AbbottAJL848,AbbottPRX}.  In addition to the binary system, isolated NSs can also emit GWs by non-axisymmetric deformations. The parameters of the quasi-normal modes (QNM) 
are dependent on the stellar interior. Therefore, knowing how QNM parameters behave in relation to the interior of NS can aid in revealing the interior of NS from QNM observations in the future. The properties of the different  oscillation modes of hybrid stars with phase transitions have been the subject of extensive research ~\citep{Flores_2014,Sandoval_2018,Rodriguez,Sandoval2022b,Zhao2022,Zhao2022b,Constantinou2021,Jaikumar2021,Kumar_2023,Constantinou2023,Kumar_2023a,Sandoval2023fmode}.  It has been noted that the quadrupolar $f$-mode, one of the QNMs, is among the most promising sources of GW emission and is also accessible with the improving sensitivity of the current LIGO-VIRGO detectors or with the next-generation GW detectors, such as the Einstein Telescope (ET) and Cosmic Explorer (CE) ~\citep{Ho2020,Pradhan2023b}. Alternatively, asteroseismology is another approach, where EOS independent or universal relations (URs)~\citep{Yagi:2013bca,Largani:2021hjo} are employed to infer the NS interior from the detection of QNM parameters~\citep{Volkel2021,Volkel2022,Pradhan2023b,Sandoval2022}. According to a recent study from ~\cite{Sandoval2023fmode}, the inclusion of a slow phase transition ( the conversion speed is slower compared to the radial perturbation time scale ) appears to violate the conventional universal relations involving $f$-mode parameters and other NS observables. Additionally, the recent study by ~\cite{Laskos2023} examines the r-mode instability windows and spin-down evolution of twin stars. It suggests that hybrid equations of state, depending on the transition density, could explain the observed spin and temperature evolution, and the future detection of r-mode GW could serve as an indicator of the presence of twin stars.
\\

In this work, we investigate the possibility of distinguishing between a sharp and slow hadron-quark phase transition via hypothetical future observations of gravitational waves from $f$-mode oscillations in NSs. Using the methodology developed in \cite{Abgaryan2018}, we parametrize the transition from hadronic to quark matter EOS beyond the Maxwell point using a single parameter, the pressure increment at the critical chemical potential. We study the effect of the nature of the phase transition on $f$-mode characteristics. Further, we examine whether detection of gravitational waves from current or future $f$-mode observations can constrain the nature of phase transition convincingly.
\\

The paper is organized in the following way. In \cref{sec:methods}, we discuss the model applied to describe the NS interior as well as its global structure. The results of our study are presented in \cref{sec:results} and we summarize our conclusions in \cref{sec:discussions}.


\section{Methods}
\label{sec:methods}
\subsection{EOS Model}\label{sec:EOS}

 The pressure density relationship, referred to as the equation of state (EOS), is vital in connecting the microscopic behavior of NS matter (NSM) to the NS observables.  As suggested in the literature,  a deconfined quark phase could appear in high-density NSM. The nature of the phase transition from hadronic to deconfined quark phase is still a matter of debate, and it is unclear whether there exists a jump in the thermodynamic variables or whether the transition is a smooth crossover. A significant number of works have described the transition either by the Maxwell or Gibbs construction.  Furthermore, consideration of finite surface tension effects can lead to the existence of a mixed phase with different geometric shapes (referred to as the ``pasta" phase)~\citep{Ravenhall,VOSKRESENSKY2003291,Yasutake2014}. The description of the crossover-like transition from hadronic to the deconfined quark phase was first discussed in ~\cite{Masuda2013} and later adequately in ~\citep{David2017,David2017b,Abgaryan2018}.

For this study, we consider the four-parameter realization of \citep{David2017,Paschalidis2018} abbreviated as ``ACB4" to describe the  NSM at densities higher than saturation density $n_0=0.15\rm fm^{-3}$, for the reasons outlined below. The ACB4 EOS model satisfies the Seidov condition at the phase transition to produce high-mass twin stars~\citep{Seidov1971}. The EOS model mimics the pasta phases and is relatively simpler to construct as required for the NS physics. The pressure ($P$) as a function of density ($n$) can be expressed as;
 \begin{align}\label{eqn:EOS_pofn}
 P(n)=\kappa_i\l(\frac{n}{n_0}\r)^{\Gamma_i},\ n_i< n< n_{i+1},\ i=1,..4, 
\end{align}

where each density region is described by the pre-factor $\kappa_i$ and polytropic index $\Gamma_i$. It is convenient and thermodynamically consistent  to express the pressure as a function of chemical potential ($\mu$) as,
 \begin{align}\label{eqn:pofmu}
 P(\mu)=\kappa_i\l[ (\mu-m_{0,i})\frac{\Gamma_i-1}{\kappa_i \Gamma_i}\r]^{\frac{\Gamma_i}{(\Gamma_i-1)}}
\end{align}
 where $m_{0,i}$ is the effective mass of the constituent: nucleon effective mass for the hadronic region and quark effective mass in the quark phase. Further, using the thermodynamic relations, the number density and energy density ($\epsilon$) can be obtained as a function of $\mu$ following the methodology given in  ~\citep{David2017,David2017b}. 
\begin{table}
    \centering
    \begin{tabular}{c| c |c |c| c}
    \hline \hline
         $i$&$\Gamma_i$& $\kappa_i$&$n_i$ &$m_{0,i}$\\
         & &$\l[\rm MeV\  fm^{-3}\r]$ & $\rm \l[fm^{-3}\r] $ & $\l[\rm MeV\r]$\\
         
         \hline
         1&4.921 & 2.1680 & 0.1650 & 939.56\\
         2& 0.0 & 63.178 & 0.3174 & 939.56\\
         3&4.00 & 0.5075 & 0.5344 & 1031.2\\
         4& 2.80 & 3.2401 & 0.7500 & 958.55\\
         \hline\hline
    \end{tabular}
    \caption{The values of the polytropic index $\Gamma_i$, the pre factor $\kappa_i$ and the constituent effective mass ($m_{0,i}$) for each 
 density region $n<n_{i}$ as given in ~\cref{eqn:EOS_pofn,eqn:pofmu} corresponding to the ACB4 parametrization from ~\citep{David2017,Paschalidis2018}. }
    \label{tab:EOS_parameter}
\end{table}
 The EOS model describes the hadron quark phase transition by the Maxwell construction with a constant transition pressure $P_c=\kappa_2$ at a critical chemical potential ($\mu_c$), making the pressure for both phases equal, i.e., $P_{H}(\mu_c)=P_{Q}(\mu_c)=P_c$. The second region with polytrope $\Gamma_2=0$ describes the first-order phase transition. In the literature, there are two different main approaches (i) replaced interpolation method (RIM) and (ii) mixed interpolation method (MIM) that have been used to mimic the pasta phase and construct the mixed phase~\cite{Abgaryan2018}. We consider the  RIM approach and replace the pressure in the relevant region of the hadronic and quark phase near the Maxwell critical point ($\mu_c,P_c$) by a parabolic polynomial function defined as ~\citep{Ayriyan,AyriyanEPJ2018};

 \begin{align}\label{eqn:pmix}
     P_{M}(\mu)=\alpha_2(\mu-\mu_c)^2+\alpha_1(\mu-\mu_c)+(1+\Delta p)P_c
 \end{align}
where $\Delta P=\Delta p*P_{c}$ represents the additional pressure of the mixed phase with $P_{c}=P(\mu_{c})$ being the critical pressure of the Maxwell construction. The mixed phase polynomial ~\eqref{eqn:pmix} connects the EOS smoothly from the hadronic phase to the mixed phase at  $\mu_H$  and mixed phase to quark phase at $\mu_q$. The unknowns $\alpha_1$, $\alpha_2$, $\mu_H$, $\mu_Q$ are determined by the continuity of thermodynamic quantities $n$ and $P$ at $\mu_H$ and $\mu_Q$ given as;
\begin{align}
    P_H(\mu_H)=P_M (\mu_H),\  P_Q(\mu_Q)=P_M(\mu_Q)  \nonumber \\
    n_H(\mu_H)=n_M (\mu_H),\  n_Q(\mu_Q)=n_M(\mu_Q)
\end{align}

\begin{figure*}
\centering 

\subfloat[]{%
  \includegraphics[width=0.5\textwidth]{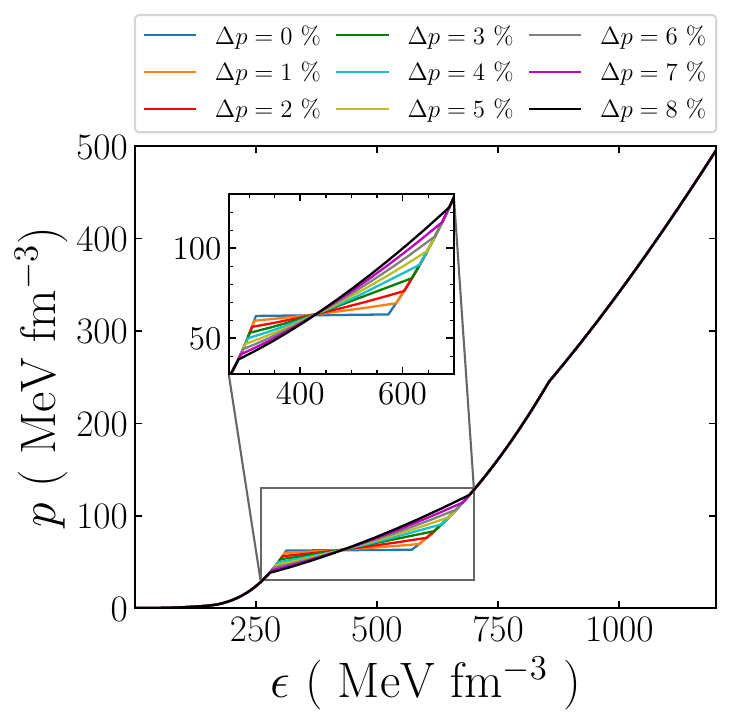}%
  \label{subfig:EOS}%
}
\subfloat[]{%
  \includegraphics[width=0.5\textwidth]{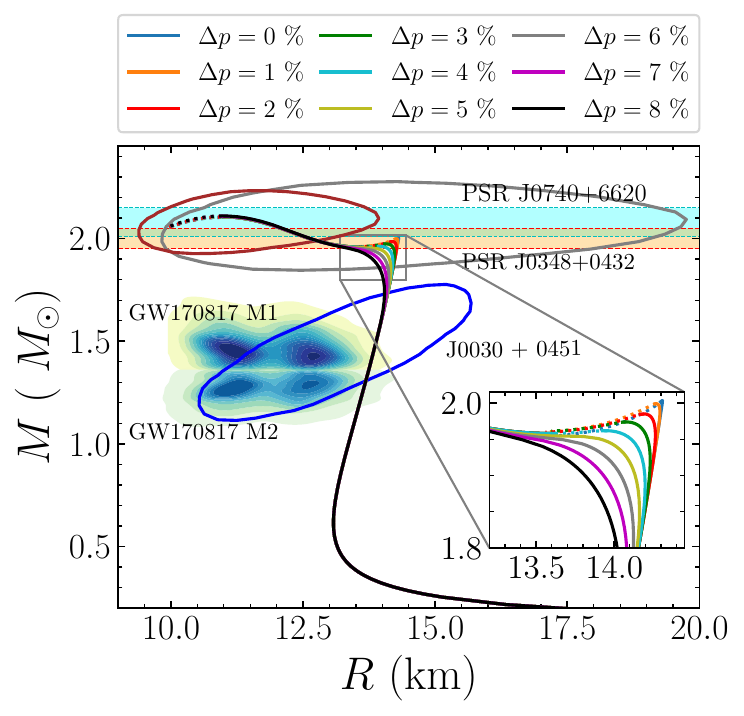}%
  \label{subfig:mr}%
}
\caption{We display (a) hybrid EOSs corresponding to ACB4 parametrization with different $\Delta p$ values (b) the corresponding $M-R$ relations. The dotted lines in ~\ref{subfig:mr} and ~\ref{subfig:mlam} show the unstable region with $\df{M}{\epsilon_c}\leq 0$. Horizontal bands correspond to masses $M=2.072^{+0.067}_{-0.066} M_{\odot}$ of PSR J0740$+$6620  and $M=2.01^{+0.04}_{-0.04} M_{\odot}$ of PSR J0348$+$0432. In ~\cref{subfig:mr},  the 90\% contours for PSR J0740+6620 corresponding to Miller et al., 2021~\citep{Miller_2021} is shown in grey color while the contour of $M-R$ measurement for PSR J0740+6620 corresponding to Riley et al., 2021 ~\citep{Riley2021} is shown in brown.  The $M-R$ estimates of the two companion neutron stars of the merger event GW170817 are shown by the shaded area labeled with GW170817 M1 (M2) in (b).}
\label{fig:EOS_MR}
\end{figure*}

We tabulate the EOS model parameters corresponding to  the ACB4 realization  in ~\cref{tab:EOS_parameter}. We display the EOSs resulting from the described  model  in ~\cref{subfig:EOS}  characterized by $\Delta p$ arbitrarily ranging from 0 up to 8\%, where $\Delta p=0$ represents the first-order Maxwell construction.  We observe that for $\Delta p$ values above 5\% the third family branch disappears implying that all the compact stars lie in a connected, second family branch. This result clearly shows how robust the existence of the third family of compact stars is within this EoS framework against the appearance of pasta phases.

\subsection{Macroscopic Observables}\label{sec:TOV}
For a given EOS $p=p(\epsilon$), the compact star (CS) configurations are obtained by integrating the general relativistic hydrostatic equilibrium Tolman–Oppenheimer–Volkoff (TOV) equations~\citep{Tolman,Oppenheimer},
\begin{eqnarray}
    \frac{dm(r)}{dr} &=& 4\pi \epsilon(r)r^{2}, \nonumber\\
    \frac{dp(r)}{dr} &=& -\frac{[p(r) + \epsilon(r)][m(r) + 4\pi r^{3}p(r)]}{r(r - 2m(r))}~.
    \label{eq:MR_relation}
\end{eqnarray}

Integrating  TOV Eqs~\eqref{eq:MR_relation}  from the center ($r=0$, with a central pressure $p=p_c$ ) to the surface of the star with the boundary condition that the pressure vanishes at the surface, i.e, $p(R)=0$, one can obtain the radius ( $R$)  of the star. The  mass enclosed within $R$ is the stellar mass ($M)$ i.e., $M=m(R)$.\\

In a binary system, the CSs get tidally deformed  due to the tidal field of the companions, and the deformation can be measured from the GW observation of the binary system represented by the  tidal deformability. The  tidal deformability can be used to constrain the CS EOS as it strongly depends upon the CS EOS. The dimensionless tidal deformability  ($\Lambda$) is expressed in terms of the love number $k_2$ in ~\eqref{eq:diml_tidaldef}. Furthermore, one needs to solve an additional set of differential equations along with the TOV equations ~\cite{Hinderer}  to obtain the love number $k_{2}$,

\begin{eqnarray}
    \Lambda= \frac{2}{3}k_{2}C^{-5}
    \label{eq:diml_tidaldef}
\end{eqnarray}
where $C$ is the compactness i.e., the ratio of mass and radius $C = M/R $.

\begin{figure}
    \centering
    \includegraphics[width=\linewidth]{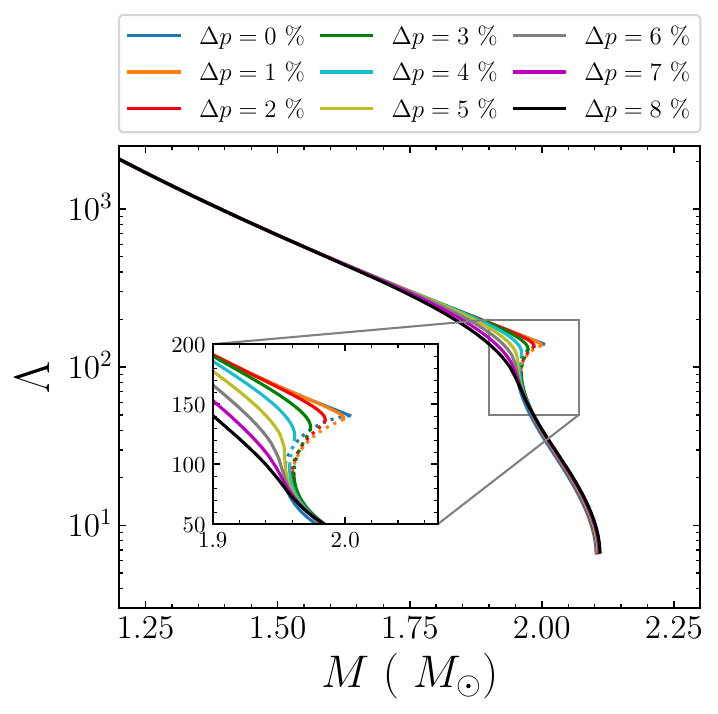}
    \caption{$M-\Lambda$  relations corresponding to the EOSs displayed in ~\cref{subfig:EOS}.}
    \label{subfig:mlam}
\end{figure}

We display the hybrid EoSs in~\cref{subfig:EOS} and the corresponding $M-R$ and $M-\Lambda$ relations in  ~\cref{subfig:mr,subfig:mlam} respectively. The second and third family branches are connected using the dotted lines, showing the unstable configurations for $\df{M}{\epsilon_c} \leq 0$.  On increasing the value of $\Delta p$, the unstable region gets shortened, and beyond $\Delta p=5\%$, the second and third families merge to form a single branch. The discontinuity in $R$ and $\Lambda$ as a function of $M$ indicates the presence of twin stars. Hence, the simultaneous inference of  $M-R$ or $M-\Lambda$ from observations can reveal the existence of discontinuity in $M-R$ or $M-\Lambda$ relations and can provide hints to the presence of twin stars or even the nature of phase transition.  

\subsection{Solving for $f$-mode Characteristics}
\label{sec:fmode_methods}
Among the different quasi-normal modes of the NS, the nonradial fundamental mode ($f$-modes) is strongly coupled with the NS fluid ~\citep{Thorne} and the dominant source of GW emission. In the literature, a significant amount of effort has been spent in developing the methodology to solve for mode characteristics including the resonance matching method~\citep{Chandrasekhar:1991}, direct integration method~\citep{Detweiler83,Detweiler85}, method of continued fraction~\citep{Leins1993,Sotani2001} and WKB approximation~\citep{Andersson96}. Though several works used the Cowling approximation method to find mode frequency ignoring the metric perturbation, the importance of inclusion of linearized general relativistic treatment over the  Cowling approximation has been discussed in several recent works ~\citep{Yoshida,Chirenti2015,Pradhan2022} which concluded that the Cowling approximation could overestimate the $f$-mode frequency upto $\sim~30\%$ compared to the frequency obtained within general relativistic treatment. 

In this work, we obtain the mode parameters by solving the perturbations in full general relativistic treatment.
 We use the direct integration method developed in ~\citep{Detweiler85,Sotani2001,Pradhan2022} to find the NS $f$-mode frequency. Shortly, the coupled perturbation equations for perturbed metric and fluid variables are integrated within the NS interior with appropriate boundary and junction (for hybrid stars with density discontinuity)  conditions ~\citep{Sotani2001}. Further, the fluid variables are set to zero outside the star, and then Zerilli's wave equation ~\citep{Zerilli} is integrated too far from the star. Further, a search is carried out for the complex $f$-mode frequency ($\omega=2\pi f+\frac{i}{\tau_f}$) corresponding to only outgoing wave solution to the  Zerilli's equation at infinity. The real part of $\omega$ represents the $f$-mode angular frequency, and the imaginary part represents the damping time. For finding the mode characteristics, we use the numerical methods developed in our previous work~\citep{Pradhan2022} along with including jump conditions for hybrid stars from~\cite{Sotani2001}.

\subsection{$f$-mode parameters and Universal Relations}\label{sec:URs}
\begin{figure*}
\centering 

\subfloat[]{%
  \includegraphics[width=0.5\textwidth]{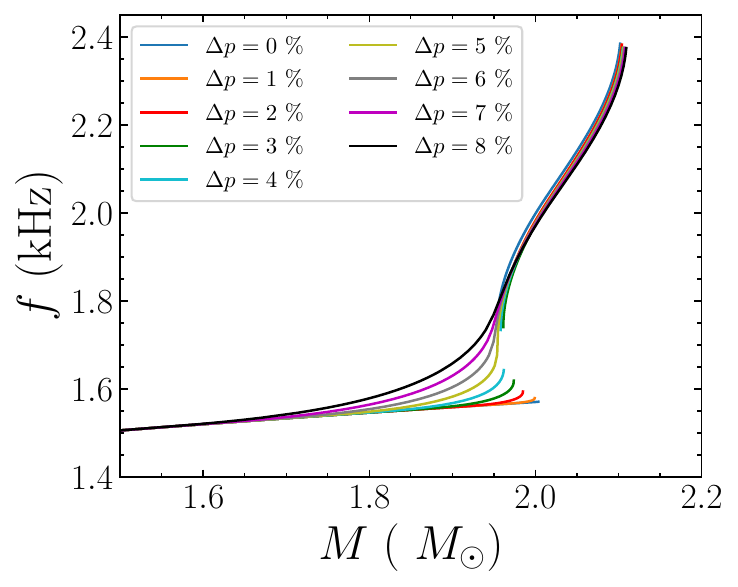}%
  \label{subfig:f_m}%
}
\subfloat[]{%
  \includegraphics[width=0.5\textwidth]{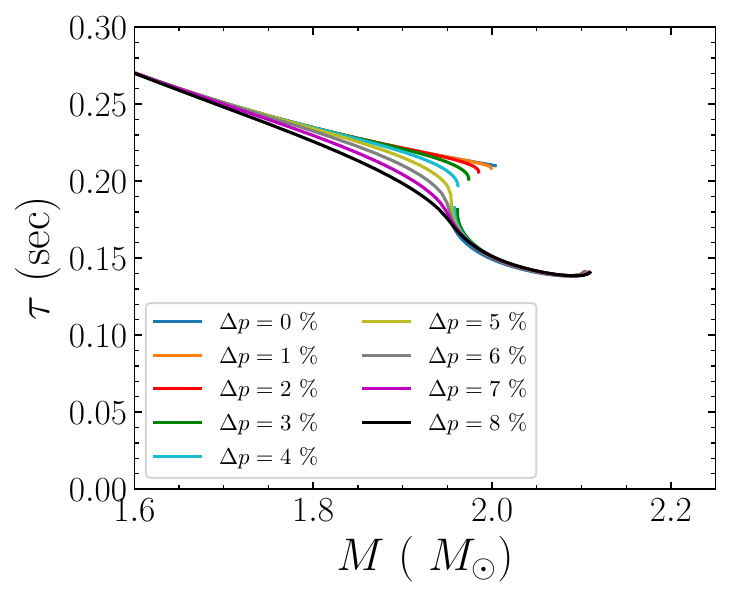}%
  \label{subfig:m_tau}%
}\qquad
\caption{We show the variation of $f$-mode (a) frequency and (b) damping time as a function of stellar mass corresponding to stable $M-R$ configurations of the EOSs presented in ~\cref{subfig:EOS}.} 
\label{fig:ftau_m}
\end{figure*}

We display in ~\cref{subfig:f_m} and~\cref{subfig:m_tau} the fundamental mode ($f$-mode) frequency ($f$) and damping time ($\tau$) respectively as a function of mass (of the stable branch) corresponding to the EOSs in~\cref{subfig:EOS}. From~\cref{fig:ftau_m}, one can conclude that there is a sudden increase in the frequency (decrease in the damping time) at the onset of the transition. However, the frequency jump disappears beyond $\Delta p=5\%$. Therefore,  simultaneous observations of $f$-mode frequency (or damping time) with mass can lead us to comment on the presence of the jump or the nature of phase transition. The simultaneous mass and mode frequency measurements can be obtained from the detection of GW events from a binary system~\citep{Williams2022}. We discuss different possibilities and prospects of binary systems involving twin stars in~\cref{sec:twinstar_binary}. However, in the case of isolated stars, we may not have the privilege of measuring the mass. The QNMs are the only source of GW emission, implying the detectable parameters are $f$ and $\tau$. Hence, one has to rely on asteroseismology to infer the stellar properties or, inversely, to infer the interior of the compact star.

\begin{figure*}
\centering 

\subfloat[]{%
  \includegraphics[width=0.5\textwidth]{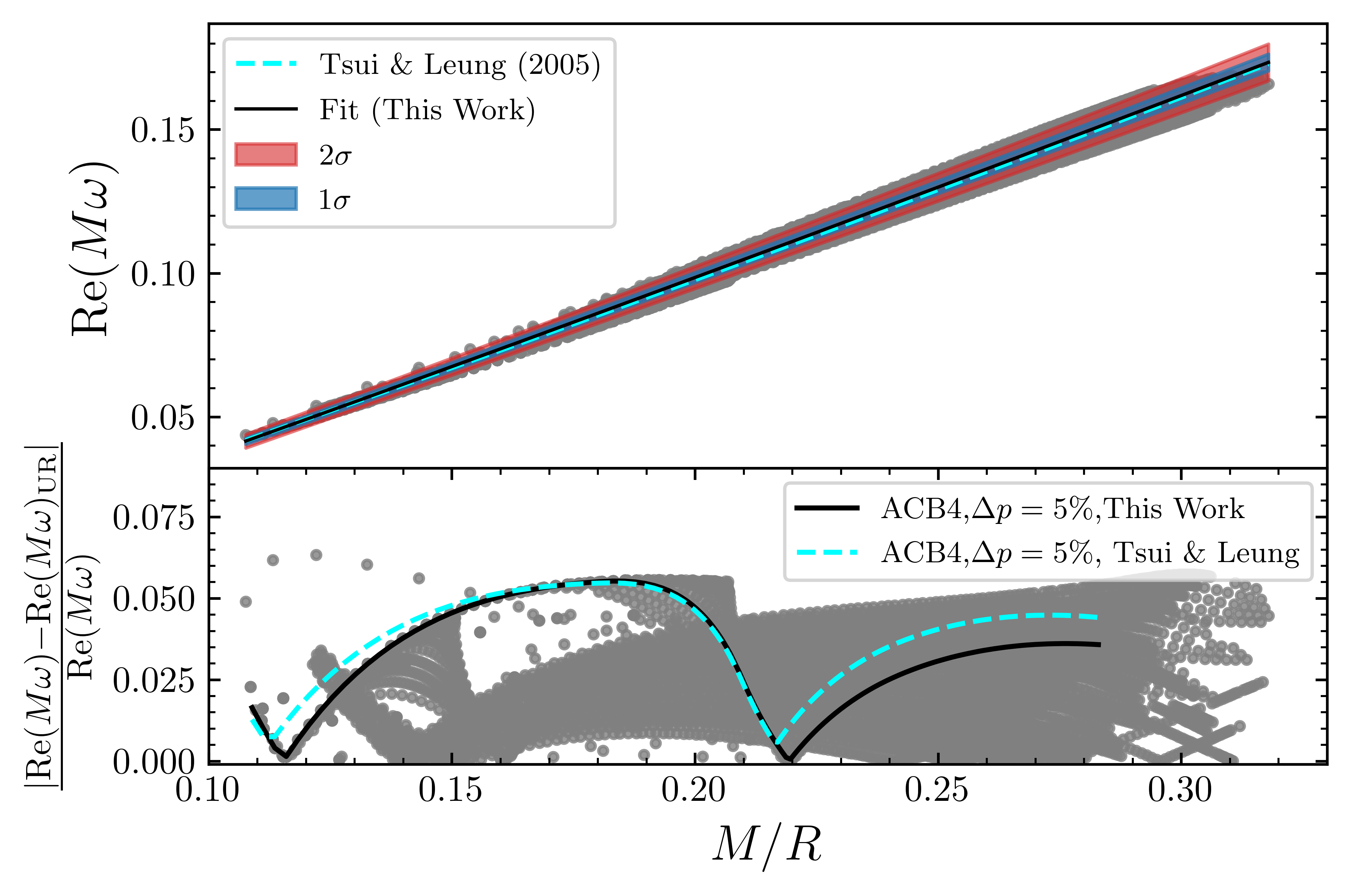}%
  \label{subfig:remw_new}%
  
}
\subfloat[]{%
  \includegraphics[width=0.5\textwidth]{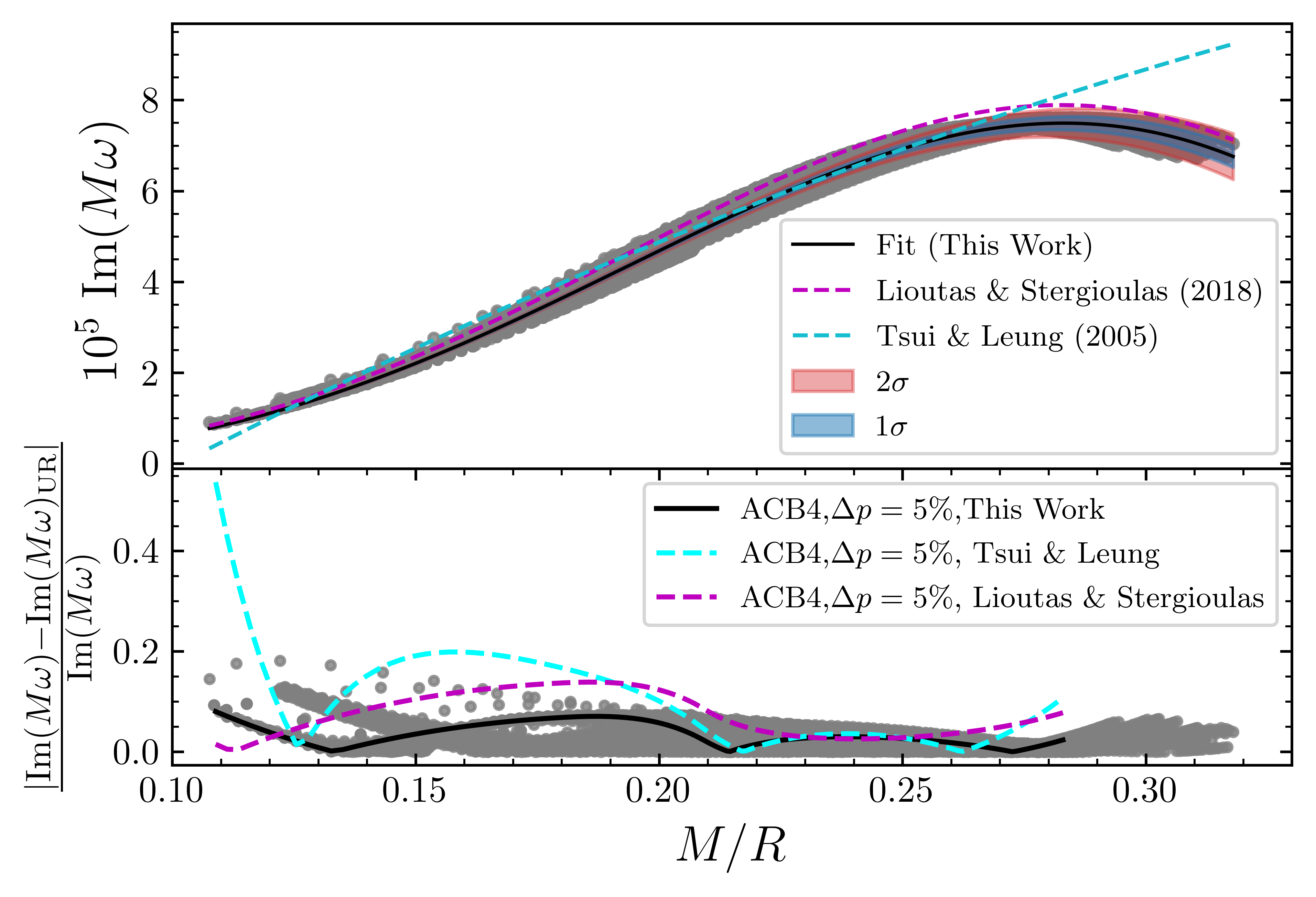}%
  \label{subfig:immw_new}%
}\qquad
\caption{Displays the  universality between  stellar compactness $M/R$ with  (a) Re($M\omega$) and (b) Im($M\omega$). The $1\sigma$ and $2\sigma$ uncertainties on the fit relations are also displayed. In both the subfigures, the lower panel displays the relative error on the $M\omega$ for all the hybrid EOSs corresponding to the fit relation ~\cref{eqn:remomega_compactness,eqn:immomega_compactness}. Additionally,  different URs and the resulting error due to different URs on the $M\omega$ of a representative hybrid EOS ACB4,$\Delta p=5\%$, are also shown. }
\label{fig:MR_UR_new}
\end{figure*}

The idea of asteroseismology involving the inference of the stellar observables from detection of  NS quasi-normal modes (QNMs) was first addressed in ~\cite{Andersson96,Andersson98}, where it was shown that there exist EOS-independent relations or universal relations (URs) among the NS mode parameters and the NS observables like $M,\ R$. Though the initial empirical relations involve the mean density of the star, later works have shown that the empirical relations may depend upon the EOS model considered~\citep{Pradhan2021,Pradhan2022,Lopez2022}. 
However, the URs involving stellar compactness ($M/R$) are EOS independent and can be used to reconstruct NS observables from the detection of QNMs~\citep{Tsui2005,Lioutas2018}.  We obtain the URs considering a wide range of EOS models considered in ~\cite{Pradhan2023}, a few hybrid EOS models from ~\cite{Ayriyan2021b,Paschalidis2018,David2021}, and the EOSs shown in ~\cref{subfig:EOS,subfig:EOS_ACB5}. All the hybrid EOS models considered in this work are displayed in ~\cref{fig:MR_new} of ~\cref{app:app_B} The URs involving stellar compactness ($C=M/R$) and mass-scaled $f$-mode angular frequency  ($M \omega$) are given in~\cref{eqn:remomega_compactness,eqn:immomega_compactness}. We display URs involving compactness, $f$-mode characteristics, and fit relations   in ~\cref{fig:MR_UR_new} and tabulate the fit parameters of~\cref{eqn:remomega_compactness,eqn:immomega_compactness} in~\cref{tab:momega_universalrelations}.

\begin{equation}\label{eqn:remomega_compactness}
    \mathrm{Re}(M\omega)=a_0+a_1\  \l(\frac{M}{R}\r)+a_2\ \l(\frac{M}{R}\r)^2
\end{equation}
 and for the UR involving  mass-scaled damping time ($M/\tau$) or Im($M\omega$) and compactness ($M/R$) can be given as,
 \begin{equation}\label{eqn:immomega_compactness}
    \mathrm{Im}(M\omega)=b_0 \  \left(\frac{M}{R}\right)^4+b_1 \ \l(\frac{M}{R}\r)^5 +b_2 \ \l(\frac{M}{R}\r)^6~.
\end{equation}

\begin{table}

    \centering
    
    \begin{tabular}{|p{0.1\linewidth}|p{0.3\linewidth}|p{0.1\linewidth}|p{0.3\linewidth}|}
    \hline
 \multicolumn{2}{|c|}{$\rm{Re}(M\omega)$}& \multicolumn{2}{|c|}{$\rm{Im}(M\omega)$} \\
 \hline
         $a_0$&-0.024$\pm 7\times 10^{-4}$  &$b_0$& $\l(10.39\pm 0.005\r) \times 10^{-2} $ \\
         \hline
         $a_1$& 0.589 $\pm$ 0.006&$b_1$& $\l(-4.87\pm 0.004\r)\times 10^{-1}$\\
         \hline
         $a_2$&0.089$\pm 0.011$ &$b_2$&$\l(5.686\pm 0.0415\r)\times 10^{-1}$\\
         \hline
    \end{tabular}
    \caption{Fit parameters for the URs \eqref{eqn:remomega_compactness} and \eqref{eqn:immomega_compactness} obtained in this work. We provide the quadratic fit for $\rm{Re}(M\omega)$ and a fit relation \eqref{eqn:immomega_compactness} for $\rm{Im} (M\omega)$.}
    \label{tab:momega_universalrelations}
\end{table}

{We present the dependence of scaled complex QNM frequency (scaled with NS mass) as a function of compactness along with the Universal Relations (URs) from this work and previous studies in  \cref{fig:MR_UR_new}. We display the URs proposed in the literature, notably those from ~\citep{Tsui2005} and ~\cite{Lioutas2018}. From \cref{fig:MR_UR_new}, it is evident that different URs are in good agreement. Additionally, we show the errors on the $f$-mode characteristics resulting from the proposed URs for the hybrid Equation of State (EOS) considered, as well as for a representative hybrid EOS, ACB4 with $\Delta p = 5\%$. We observe better agreement for Re($M\omega$) among the proposed URs and those given in~\cite{Tsui2005}. The deviation of the fit relation of ~\cite{Lioutas2018} from General Relativity (GR) can be attributed to the approximations considered for solving the $f$-mode damping times in their work. For the Im$(M\omega)$, we notice a comparatively lower error with the proposed URs compared to the relation given in  \cite{Tsui2005}. For the URs involving the love number, e.g., the $f-\mathrm{Love}$ relation discussed in the ~\cref{app:URs_add},  although the different URs introduce different uncertainties in different ranges of $\Lambda$,  we notice that the errors are minimal, irrespective of the UR used. As we focus on the distinguishability of hybrid stars and to avoid clutter/confusion in \cref{fig:recovery_m_r_compur} and  \cref{fig:recovery_m_r_mwlambdaur}, we only show the estimations of Mass-Radius ($M-R$) recovered with the newly obtained URs. The uncertainties in the $M-R$ estimation from future $f$-mode observations discussed later in \cref{sec:detection} result from the measurement uncertainties on the mode frequency and damping time. Therefore, fixing the URs to a different one will not change the uncertainties in the $M-R$ measurement, as discussed in \cref{fig:mr_recovery_ET_URM2}. However, a change in UR might lead to a bias or shift in the $M-R$ measurement.}

Recently, a new class of  URs has been proposed in~\cite{Sotani2021}, involving the tidal deformability ($\Lambda$) scaled $f$-mode parameters and later considered in~\cite{Sandoval2022} for $wI$-modes. The new URs involving $\Lambda$ can be written as,

\begin{equation}\label{eqn:wlambda_mlambda}
    \log \l[\frac{\omega_{R,I}}{\Lambda}\r] = \alpha_{R,I} + \beta_{R,I} x +\gamma x^2
\end{equation}
where, $x=\log\l[\Lambda M_{1.4M_{\odot}}\r]$ with $M_{1.4M_{\odot}}=M/1.4M_{\odot}$, $\omega_R=\rm{Re}(\omega)=2\pi f$ and $\omega_I=\rm{Im}(\omega)=1/\tau$ . Similarly, other URs exist for  $\frac{\omega_{R,I}}{\Lambda}$ with radius scaled tidal deformability parameter $\Lambda R_{10\rm km}$ and can be written as,
\begin{equation}\label{eqn:wlambda_radlambda}
    \log \l[\frac{\omega_{R,I}}{\Lambda}\r] = \alpha^{\prime}_{R,I} + \beta^{\prime}_{R,I} y +\gamma^{\prime} y^2
\end{equation}
where, $y=\log\l[\Lambda R_{10}\r]$ with $R_{10}=R/{10 \rm km }$.

Hence, there will be 2 URs for \cref{eqn:wlambda_mlambda}: one for $f$-mode angular frequency Re($\omega)$ and the other for the damping time, and similarly 2 URs for \cref{eqn:wlambda_radlambda}. The URs for $\l[\frac{\omega_{R,I}}{\Lambda}\r]$ as a function of $\ln\l[M_{1.4M_{\odot}}\Lambda\r]$   are displayed in~\cref{subfig:omega_mlambda} of ~\cref{app:URs_add} and the fit parameters are tabulated in~\cref{tab:remomeg_mlambda}. We display the URs among $\omega_{R,I}$ with $\ln\l[\Lambda R_{10\mathrm{km}}\r]$ ~\cref{subfig:omega_rlambda} of ~\cref{app:URs_add} and tabulate the fit parameters in~\cref{tab:remomeg_radlambda}.

\begin{table}

    \centering
    
    \begin{tabular}{|p{0.1\linewidth}|p{0.3\linewidth}|p{0.1\linewidth}|p{0.3\linewidth}|}
    \hline
 \multicolumn{2}{|c|}{$\log \l[\frac{\omega_R}{\Lambda}\r]=f(x)$}& \multicolumn{2}{|c|}{$\log \l[\frac{\omega_I}{\Lambda}\r]=f(x)$} \\
 \hline
         $\alpha_R$&3.32 $\pm 1.49 \times 10^{-3}$  &$\alpha_I$& 1.92 $\pm 3.4 \times 10^{-3} $ \\
         \hline
         $\beta_R$&-1.04$\pm 6.69\times10^{-5}$  &$\beta_I$& -0.64$\pm 8.9 \times 10^{-4} $\\
         \hline
         $\gamma_R$&-0.0178$\pm 6.92\times10^{-5}$  &$\gamma_I$&-0.071 $\pm 1.03 \times 10^{-4} $\\
         \hline
    \end{tabular}
    \caption{Fit parameters for real  ($\omega_R$) and imaginary ($\omega_I$)  of the complex frequency $\omega$, related to $x=\log\l[M_{1.4M_{\odot}}\Lambda\r]$ through the URs~\cref{eqn:wlambda_mlambda}. The coefficients corresponding to $\omega_R$ and $\omega_I$ are denoted with subscript $R$ and $I$, respectively.  }
    \label{tab:remomeg_mlambda}
\end{table}

\begin{table}

    \centering
    
    \begin{tabular}{|p{0.1\linewidth}|p{0.25\linewidth}|p{0.1\linewidth}|p{0.25\linewidth}|}
    \hline
 \multicolumn{2}{|c|}{$\log \l[\frac{\omega_R}{\Lambda}\r]=f(y)$}& \multicolumn{2}{|c|}{$\log \l[\frac{\omega_I}{\Lambda}\r]=f(y)$} \\
 \hline
         $\alpha^{\prime}_R$&2.82 $\pm 1.5\times 10^{-3}$  &$\alpha^{\prime}_I$&1.68 $\pm 1.08 \times 10^{-3}$ \\
         \hline
         $\beta^{\prime}_R$&-1.01$\pm 7\times10^{-4}$  &$\beta^{\prime}_I$& -0.73$\pm 8.31\times 10^{-4} $\\
         \hline
         $\gamma^{\prime}_R$&-0.003$\pm 7.28\times10^{-5}$  &$\gamma^{\prime}_I$& -0.041$\pm 8.41 \times 10^{-5} $ \\
         \hline
    \end{tabular}
    \caption{Fit parameters for real  ($\omega_R$) and imaginary ($\omega_I$)  of the complex frequency $\omega$, related to $y=\log\l[\Lambda R_{10}\r]$ through the URs~\cref{eqn:wlambda_radlambda}. The coefficient corresponding to $\omega_R$ and $\omega_I$ are denoted with subscript $R$ and $I$ respectively.}
    \label{tab:remomeg_radlambda}
\end{table}

{The universal relations (URs) discussed in this study incorporate a broad spectrum of hybrid equations of state (EOSs) derived from several sources, including ~\cite{Ayriyan2021b,Paschalidis2018,David2021}, in addition to the EOSs employed in the investigation by ~\cite{Pradhan2023b}. Although utilizing a limited set of hybrid EOSs might yield slightly different fitting relations, primarily in the high compactness region depending on the onset point, the theoretical values for the hybrid EOSs fall well within the uncertainty bands arising from a wide range of NS EOSs with varying stiffness. However, we notice that the inclusion of a large number of hybrid EOSs introduces a greater uncertainty in the URs, particularly in the high compactness region ($M/R > 0.25$). Recent investigations indicate that the presence of a hadron-quark interface in a compact star, where the conversion speed is slow compared to the radial perturbation, can lead to an extended twin branch of slow stable hybrid stars (SSHS) in the mass-radius ($M-R$) plane~\citep{Pereira2018,Pereira_2022,Sandoval2022b,Sandoval2023fmode}. It is noteworthy that the recent work by ~\cite{Sandoval2023fmode} discusses how the inclusion of SSHS leads to deviations in the URs compared to earlier findings. However, this study omits such SSHS scenarios in deriving the URs, and this matter will be addressed in future investigations. Thus, the twin stars discussed in this study should not be confused with SSHS.}

\section{Results}\label{sec:results}

As discussed, GW asteroseismology aims to recover the stellar properties from the detected mode parameters using the URs. We discuss the impact of uncertainties associated with the URs on the reconstruction of stellar properties from $f$-mode parameters in~\cref{sec:asteroseismology}. Additionally, the detection of GWs from $f$-modes itself can be associated with uncertainty in the mode parameters, which will reflect on the stellar properties, which is discussed in~\cref{sec:detection}. Furthermore, we discuss the inverse problem to constrain the compact star EOS from $f$-mode GW observation in~\cref{sec:inverse_problem}.

\subsection{Effect of uncertainty in the URs}\label{sec:asteroseismology}
As mentioned in the ~\cref{sec:URs}, the idea behind constructing the URs is to reconstruct the NS observables like $M$, $R$ from detecting mode parameters. For an observed $f$-mode  frequency $\omega$, the URs ~\cref{eqn:remomega_compactness} will result in one curve in the $M$ and $M/R$ plane corresponding to  Re($\omega$) and similarly for the observed damping time $\tau$, the UR ~\cref{eqn:immomega_compactness} will result in another curve in the $M$ and $M/R$ plane corresponding to the Im($\omega$). The intersection of the two resulting curves will give us the value of $M$ and $M/R$, hence $R$. As one of our main concerns is to probe the nature of the phase transition, which can be decided from the value of  $\Delta p$ or focusing on the particular $M-R$ region of the ~\cref{subfig:mr}: mainly in the mass range 1.8$M_{\odot}$ to 2.0 $M_{\odot}$ for the EOS model considered here. Hence, if the observed $M$ and $R$ along with their uncertainties of one EOS $\Delta p\neq 0\%$ do not overlap with the  unstable region of the $M-R$ curve of the $\Delta p= 0$, then that could lead us to confirm that the $f$-mode observations can differentiate the nature of $\Delta p$. So, assuming that the $f$-mode parameters are observed precisely for a few randomly selected stars,  the $M$ and $R$ can be estimated using URs, as explained before. The consideration of the uncertainties associated with the UR parameters will further result in uncertainties of the recovered  stellar observables such as $M$ and $R$.

\begin{figure*}
\centering
    \subfloat[]{%
  \includegraphics[width=0.45\linewidth]{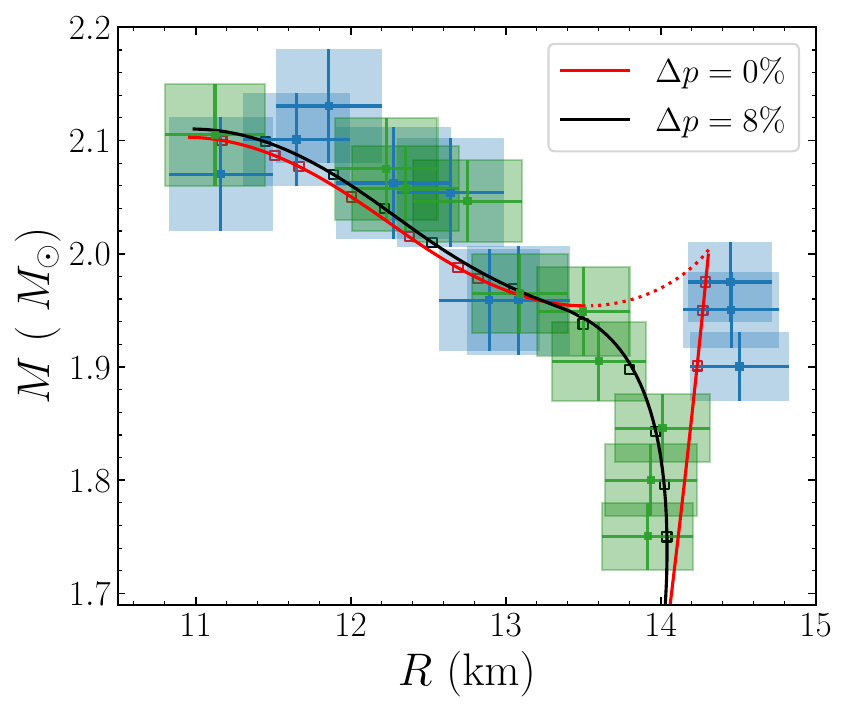}
   \label{fig:recovery_compur_dp0and8}
    }
    \subfloat[]{%
      \includegraphics[width=0.45\linewidth]{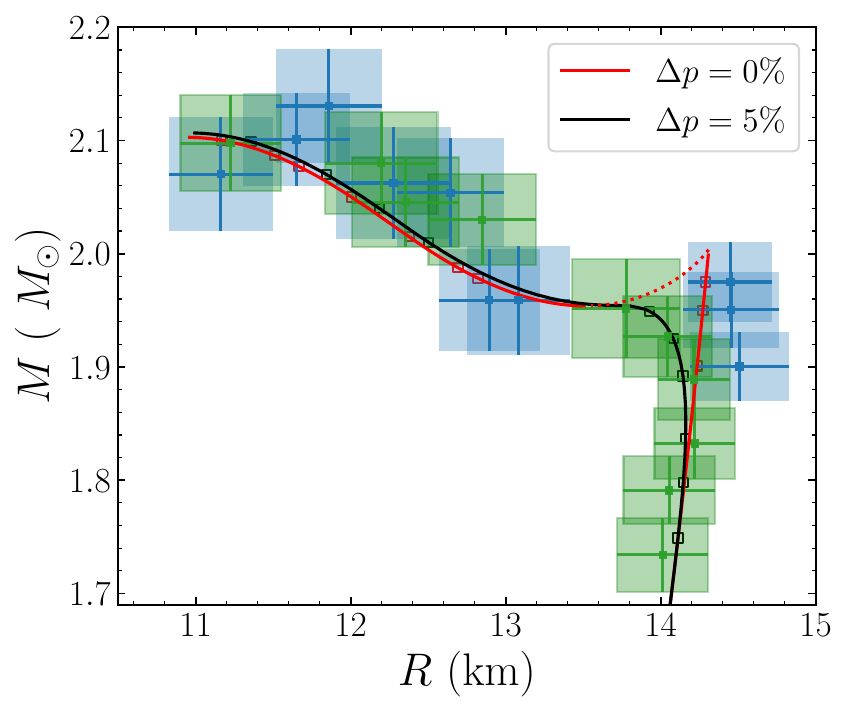}
  \label{fig:recovery_compur_dp0and5}
   }
    \caption{(\ref{fig:recovery_compur_dp0and8}) Recovered $M$ and $R$ using URs from ~\cref{eqn:remomega_compactness,eqn:immomega_compactness}. The assumed configurations are shown with empty  red squares (for $\Delta p=0\%$) and black squares (for $\Delta p=8\%$). The uncertainty regions of the recovered $M$ and $R$ are shaded in blue for $\Delta p=0\%$ (green for $\Delta p=8\%$). (\ref{fig:recovery_compur_dp0and5}) Same as \ref{fig:recovery_compur_dp0and8}, but the EOSs considered to be differentiate are $\Delta p=0\%$ and $\Delta p=5\%$~.} 
    \label{fig:recovery_m_r_compur}
    \end{figure*}

Under the assumed scenario of the precise measurement of the mode parameters, the mass and radius recovered for different EOSs sing URs ~\cref{eqn:remomega_compactness,eqn:immomega_compactness} with different $\Delta p$   along with their uncertainties are displayed in the ~\cref{fig:recovery_m_r_compur}. It is clear that the uncertainty region recovered for $M$ and $R$ for   $\Delta p=8\%$ are distinguishable from that of recovered for $\Delta p=0\%$, particularly in the region where the $M-R$ sequence of $\Delta p=0\%$ have an unstable region (connecting the second and third family stable branches) within it (see ~\cref{fig:recovery_compur_dp0and8}). This indicates that observing $f$-mode parameters can lead to commenting about the value of $\Delta p$ and hence the nature of phase transition. However, if one considers the EOSs with $\Delta p=5\%$, the recovered regions for $M$ and $R$ overlap with $M$-$R$ recovered  from EOS with $\Delta p=0\%$ and indicated no distinguishability. Hence although the observations of $f$-modes and the use of URs can differentiate the $M$-$R$ region for EOSs with $\Delta p=8\%$ from EOS with $\Delta p=0\%$, it fails to differentiate the $M-R$ region between  $\Delta p=0\%$ and $\Delta p=5\%$.

\begin{figure}
    \centering
    \includegraphics[width=0.95\linewidth]{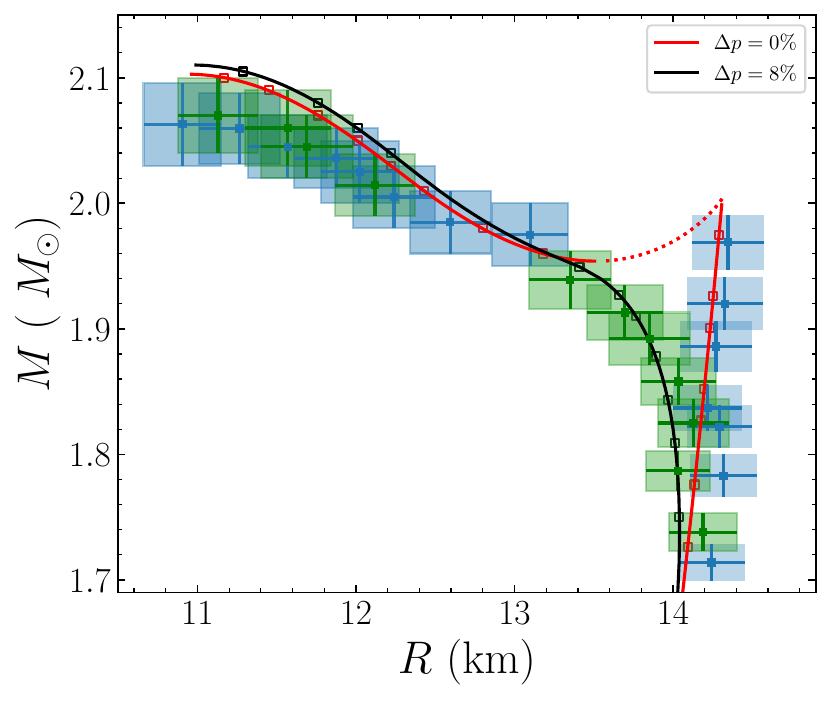}
    \caption{ Same as \cref{fig:recovery_compur_dp0and8} but the $M-R$ are recovered using the URs from ~\cref{eqn:wlambda_mlambda,eqn:wlambda_radlambda}.}
    \label{fig:recovery_m_r_mwlambdaur}
\end{figure}

The analysis can be performed using the URs from ~\cref{eqn:wlambda_mlambda,eqn:wlambda_radlambda}. The methodology of recovering stellar properties using the URs ~\cref{eqn:wlambda_mlambda,eqn:wlambda_radlambda}, have been demonstrated in ~\cite{Sandoval2022b} for axial w modes, which can be extended to the $f$-mode. 
Performing similar tests as mentioned above but with the URs \cref{eqn:wlambda_mlambda,eqn:wlambda_radlambda}, the recovered $M$ and $R$ regions for EOSs $\Delta p=0\%$ and $\Delta p=8\%$ are displayed in ~\cref{fig:recovery_m_r_mwlambdaur}. We notice that using any pair of URs results in similar conclusions regarding the distinguishability of the value of $\Delta p$. However, in the case of using URs  ~\cref{eqn:wlambda_mlambda,eqn:wlambda_radlambda}, the URs  ~\cref{eqn:wlambda_mlambda} provides the measurement of ($M$,$\Lambda$) and use of ~\cref{eqn:wlambda_radlambda} recovers  ($R$,$\Lambda$) hence, during the recovery, it is needed to be checked that the resulting $\Lambda$  with different URs should be same or should be within minimal measurement uncertainty. In principle,  to get the joint posterior of ($M, R$), one has to eliminate the $\Lambda$ from ($M$,$\Lambda$) and   ($R$,$\Lambda$) obtained using ~\cref{eqn:wlambda_mlambda} and ~\cref{eqn:wlambda_radlambda} respectively.

\subsection{Effect of Observational Uncertainties}\label{sec:detection}
~\Cref{sec:asteroseismology} demonstrates the  $f$-mode asteroseismology problem dealing with the uncertainty on the URs  under the assumption that the mode parameters are measured precisely. However, detecting $f$-mode will always result in some uncertainties in the mode parameters, which we address in this section while discussing the fact that whether or not future observation can help us distinguish the nature of phase transition. The $f$-mode GW  signal for a source at a distance $d$, with frequency ($f$) and damping time scale ($\tau$), can be modeled as a damped sinusoidal ~\citep{Kokkotas2001,Ho2020}, 
\begin{equation}\label{eqn:gwwaveform}
        h(t)=h_0e^{-(t-t_0)/\tau} \sin{\l[2\pi f (t-t_0)+\phi\r]} ,\ \mathrm{for} \ t\geq t_0
\end{equation}
where, 
\begin{equation}\label{eqn:h0}
h_{0}= 4.85  \times 10^{-17} \sqrt{\frac{E_{\rm gw}}{M_{\odot}c^2}} \sqrt{\frac{0.1 {\rm sec}}{\tau}} \frac{1 \rm{kpc}}{d}\left(\frac{1 \rm{kHz}}{f}\right)~.
\end{equation}
  
One needs the  value of the $E_{\rm gw}$ for proceeding further. There are different phenomena have been used to stimulate the excitation of the  NS oscillation modes: like mini collapse~\citep{Lin2011}, newborn NSs~\citep{Ferrari2003}, star quakes ~\citep{Keer2014,Mock_1998}, magnetars~\citep{Abbott_2019,AbbottLVK2022},  the pre-merger ~\citep{Steinhoff2016,Andersson2018} and post-merger stages of a NS in binary~\citep{Shibata1994,Stergioulas2011,Bauswein2012}. Even though the connection between pulsar glitches and the mode excitation is not clear, the assumption of $f$-mode excitation with an energy similar to that of typical pulsar glitches has been widely considered while discussing NS seismology~\citep{Andersson2001,Andersson2021} and the detectability~\citep{Ho2020,Abbott_2019,AbbottPRD104,AbbottFRB2022,AbbottLVK2022} of transient $f$-mode GW signal.  A significant number of explorations have been made regarding the relationship between glitches and $f$-mode excitations in recent works~\citep{Keer2014,Yim2023,Lopez2022}. Recent works have also been performed  on $f$-mode GW searches with the assumption of mode excitation with energy typical to that of pulsar glitches by LIGO-VIRGO-KAGRA (LVK) collaboration ~\citep{Abbott_2019,AbbottFRB2022,AbbottLVK2022,AbbottPRD104}. Hence to consider the observational uncertainties  in this work, we consider the $f$-mode excitation in the isolated glitching pulsars with energy  same as the typical energy of pulsar glitches. Now assuming that  GW energy $E_{\rm gw}$ is supplied by the energy of the glitch; one can have ~\citep{Ho2020},
\begin{equation}\label{eqn:egw}
    E_{\text{gw}}=E_{\text{glitch}}=4\pi^2I\nu^2 (\frac{\Delta \nu}{\nu})~,
\end{equation}
where $I$ and $\nu$ are the moments of inertia and spin frequency, respectively, whereas  $\frac{\Delta \nu}{\nu}$ is the relative change in spin frequency of a glitch event. 

\begin{figure*}
\centering
    \subfloat[]{%
  \includegraphics[width=0.4\linewidth]{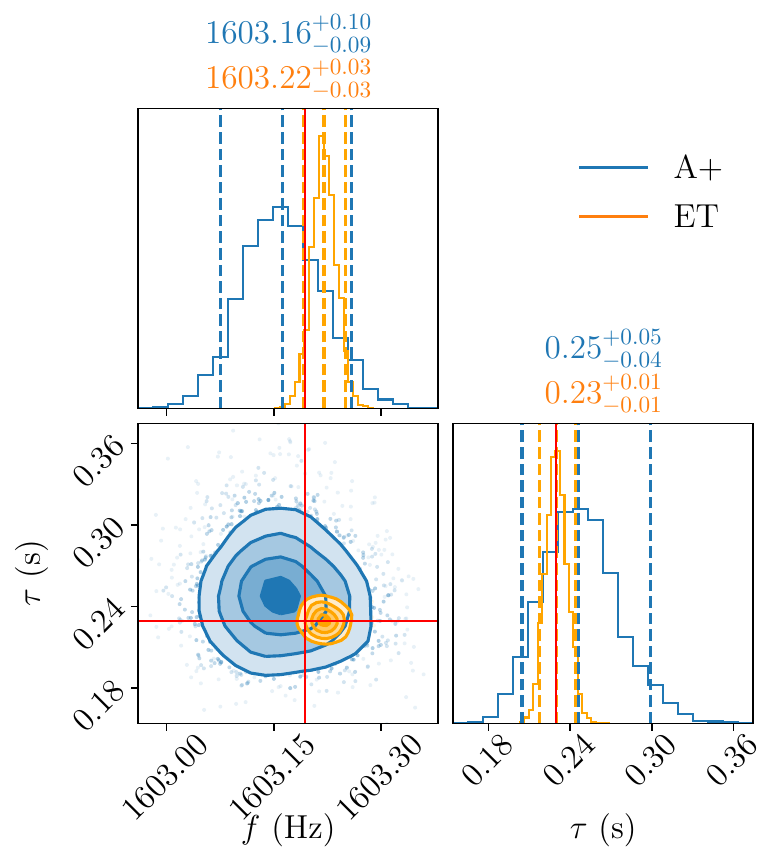}
   \label{fig:f_tau_aplus_ET}
   }
    \subfloat[]{%
      \includegraphics[width=0.4\linewidth]{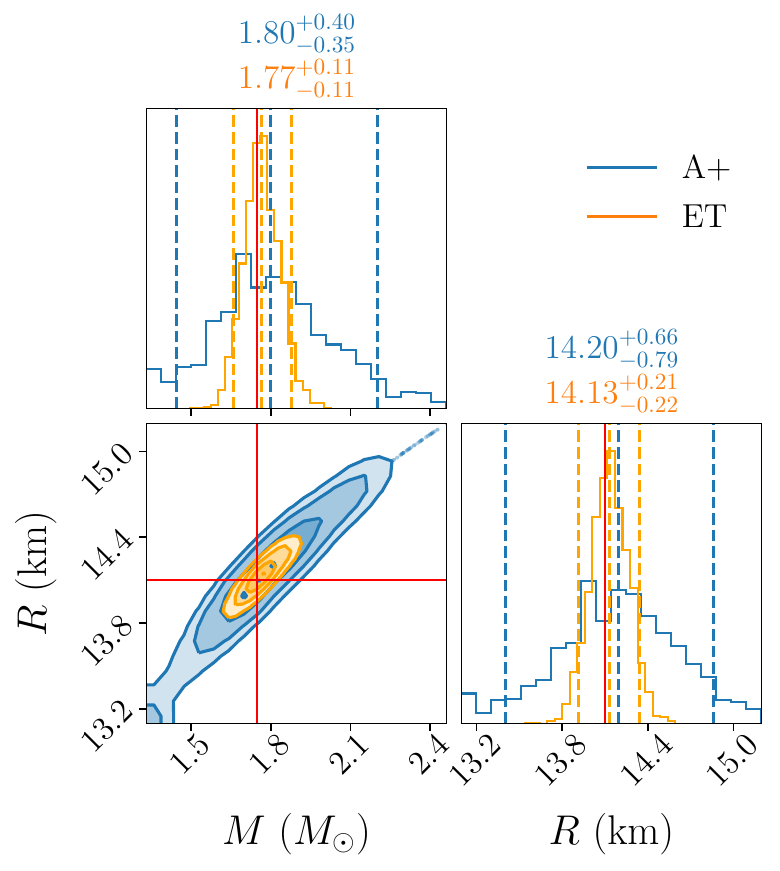}
   \label{fig:mr_aplus_ET}
   }
    \caption{(a) Marginalised corner distribution of  $f$ and $\tau$ recovered with different GW network configurations (blue for $\rm A+$ and orange for ET). Red lines mark the injected values. The injected values for $f=1603.2$ Hz and $\tau= 0.23$ s corresponds to $M=1.75M_{\odot}$  and  EOS with $\Delta p=0\%$.  (b) Figure showing the joint distribution of $M$ and $R$ obtained by using the URs ~\cref{eqn:remomega_compactness,eqn:immomega_compactness} from the   recovered ($f,\tau$) posterior. The injected values of  $M=1.75M_{\odot}$ and $R=14.10$ km are shown using the red lines. In both figures, the title shows the median and symmetric 90\% credible intervals of the  parameters.  }
    \label{fig:f_tau_and_mr_corner}
    \end{figure*}

To demonstrate the asteroseismology problem, we consider an $f$-mode GW  event from Vela pulsar with an  energy corresponding to the strongest glitch of the Vela pulsar and assign a random mass $1.75M_{\odot}$. Then the other  stellar properties required for modeling the GW signal as per ~\cref{eqn:gwwaveform,eqn:h0,eqn:egw}, such as $f,\tau, I$  corresponding  to a $1.75M_{\odot}$ star is assigned  from the  hybrid EOS model with $\Delta p=0\%$. Then we perform the parameter estimation (PE) of the GW parameters in a Bayesian framework using the  nested sampling algorithm \textit{dynesty}~\citep{dynesty}, as implemented on the GW inference package~\textit{bilby}~\citep{Bilby}. We keep  a log uniform prior in $E_{\rm gw}\in \rm logU[10^{-25},10^{-4}]  $, uniform prior in $f\in U[500,4000]\rm Hz$, uniform prior in $\tau \in \rm U[0.05,0.5] s$. We keep the distance $d$ and sky positions fixed at their observed values. We consider two GW network configurations: first, 2 LIGO detectors H1, L1 operating at O5 sensitivity~\cite{Abbott2020}~\footnote{\url{https://dcc.ligo.org/LIGO-T2000012/public}} as anticipated for the 5th observation run (A+), and then consider the  next generation Einstein telescope (ET) with ET-D sensitivity~\citep{Hild_2011}\footnote{\url{https://dcc.ligo.org/LIGO-T1500293/public}}. We display the  joint distribution of $(f,\tau)$ (marginalised over $E_{\rm gw}$) recovered with A+ and ET in ~\cref{fig:f_tau_aplus_ET}. Further, we reconstruct the ($M,R$) from the recovered posterior $(f,\tau)$ using the URs  ~\cref{eqn:remomega_compactness,eqn:immomega_compactness} and display the recovered ($M,R$) in ~\cref{fig:mr_aplus_ET}.

It is clear that in both   $\rm A+$ and ET configuration, with a 90\%
credible interval (CI), the frequency $f$ can be estimated within $\leq 1\%$. However, for the same source, the damping time $\tau$ can be recovered within $\sim 5\%$  in ET compared to the  $\sim$ 20\% error  recovered with A+ configuration. For the injected scenario in $\rm A+$, the $M$ and $R$ with a 90\%  credible interval are recovered within $\sim$20\% and $\sim 6\%$, respectively. In ET, the mass and radius, with a 90\% credible interval, are recovered within $\sim$6\%  and $\sim 2\%$, respectively.  It is not always that the posteriors of recovered ($M-R$) is unique as shown in ~\cref{fig:mr_aplus_ET}. The URs that have given in ~\cref{eqn:remomega_compactness,eqn:immomega_compactness}  are not linear, and that can result in more than one  solution in the ($M,R$) plane for a given pair of ($f,\tau$). In the case of multiple solutions, one can further combine different types of URs wisely  to find out the correct solution for ($M,R$) better representing the observational data ($f,\tau$).
Furthermore, we only use the URs~\cref{eqn:remomega_compactness,eqn:immomega_compactness} to recover the stellar observables from the posterior of ($f,\tau$)  and do not consider the URs ~\cref{eqn:wlambda_mlambda,eqn:wlambda_radlambda}, as one needs to perform additional works on eliminating the $\Lambda$ to get the joint posterior of $(M,R)$. Here, we also ignore the uncertainties on the URs as we focus on the observational uncertainties.


\subsubsection{Can we differentiate the nature of $\Delta p$ from future $f$-mode GW observations?}
\begin{figure}
    \centering
      \includegraphics[width=0.84\linewidth]{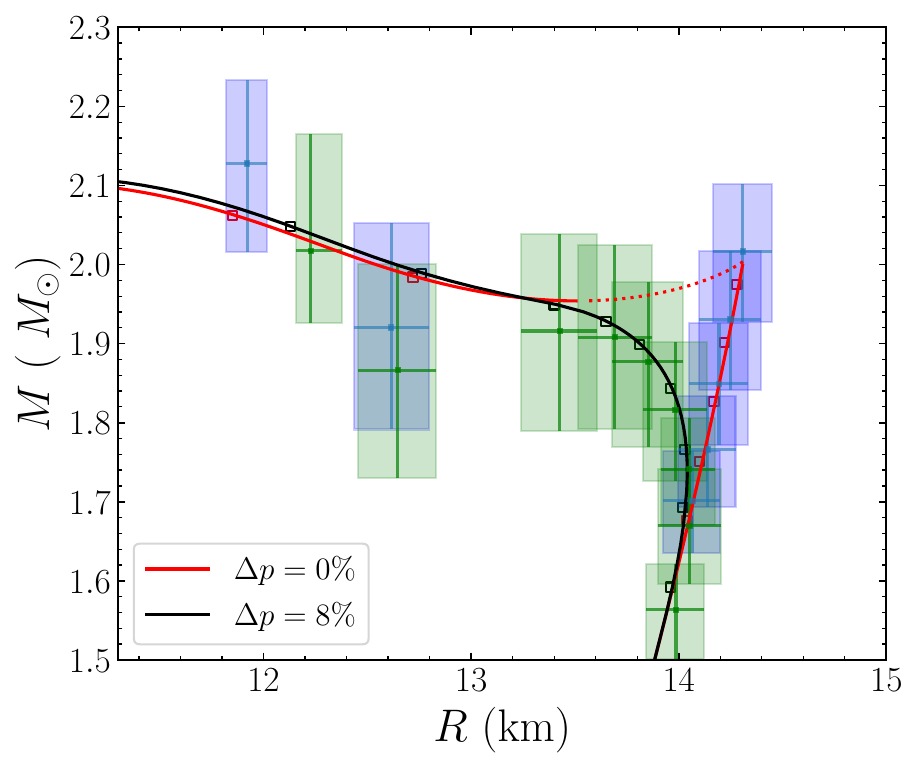}
  
       \caption{ Recovered $M-R$ using URs ~\cref{eqn:remomega_compactness,eqn:immomega_compactness} from a posterior of recovered ($f,\tau$) for a random choice of masses and glitches injected in ET. The injections are shown with empty squares (in red for $\Delta p=0\%$ and black for $\Delta p=8\%$). The uncertainties are shown in  blue (green) for EOS with $\Delta p=0\%$ ($\Delta p=8\%$). }
    \label{fig:mr_recovery_ET_URM2}
\end{figure}

  To check whether the future observations of $f$-modes can help us distinguish the nature of $\Delta p$  from the $M-R$ plane, the following tests are performed (as the uncertainties on the measurements of ($M,R$) in A+ is quite high, we investigate this with ET): 

\begin{itemize}
    \item Few $f$-mode GW events are considered with random masses with assumption of the particular EOS is $\Delta p=0\%$. A random glitch energy  is assigned to each event  such that the Signal to Noise Ratio (SNR) in ET is $\geq 10$~\footnote{The data of the glitching pulsars are taken  from the Jodrell Bank Glitch Catalogue \citep{jbglitch11}. The Jodrell Bank Glitch Catalogue lists each detected glitch's relative spin frequency change. For the pulsar's spin frequency $\nu$, distance $d$, and sky position, we use the ATNF Pulsar Catalogue \citep{atnf05}.}. We perform a parameter estimation using Bilby to get the posteriors of ($f,\tau$). From the recovered ($f,\tau$), we reconstruct the ($M,R$) using the URs ~\cref{eqn:remomega_compactness,eqn:immomega_compactness}.
    \item We repeat the above exercise considering the EOS with $\Delta p=8\%$ by choosing a few $M-R$ configurations where there is an unstable region in the $M-R$ plane for the EOS $\Delta p=0\%$.

From~\cref{fig:mr_recovery_ET_URM2}, one can conclude that there are overlapped regions of  recovered $M-R$  for different EOSs. Additionally, the uncertainty on the $M-R$ recovered for the EOS model with $\Delta p=8\%$ overlaps with the particularly unstable region connecting the second and third family of the stellar configurations in the $M-R$ relation of the $\Delta p=0\%$ EOS model, and one can barely distinguish the value of  $\Delta p$ based upon the recovered $M-R$. However, it is clear that under the assumption of $\Delta p=0\%$, i.e., in the case of the presence of twin stars, the $M-R$ of the star from the second family  and its twin companion from the third family is clearly distinguishable, which can lead to confirming the presence of twin stars. We have also considered the scenario distinguishing the EOS models with $\Delta p < 8\%$ other than 8\%  from $\Delta p=0\%$. However, one can notice that even the consideration of the extreme cases of $\Delta p$  (0\% and 8\%) in this work results in overlapping regions in the recovered  $M-R$ and further consideration of lower $\Delta p$ values makes the distinguishability of the nature of $\Delta p$ more challenging. 
\end{itemize}

\subsection{Inverse Problem: constraining the EOS parameters}\label{sec:inverse_problem}
Future observations of $f$-mode GW events can be further used to constrain the interior of the compact stars by using the observed $f$-mode parameters~\citep{Volkel2021,Volkel2022,Pradhan2023b}. As the mode parameters can be observed more precisely with the next generation GW detectors, the posterior can then be translated to get a better  constraint on  the  EOS model parameters. Furthermore, depending upon the nature and posterior of $\Delta p$, further inferences can be drawn regarding the nature of the preferred phase transition or even the existence of the pasta phase.  

\subsection{Sensitivity to Twin star detection  for low mass twins}
\begin{figure*}
\centering 

\subfloat[]{%
  \includegraphics[width=0.5\textwidth]{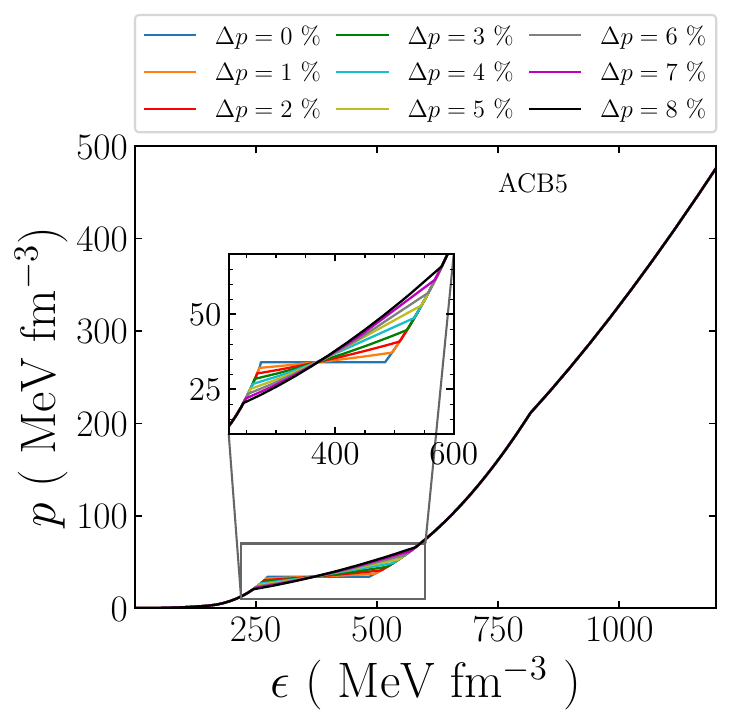}%
  \label{subfig:EOS_ACB5}%
}
\subfloat[]{%
  \includegraphics[width=0.5\textwidth]{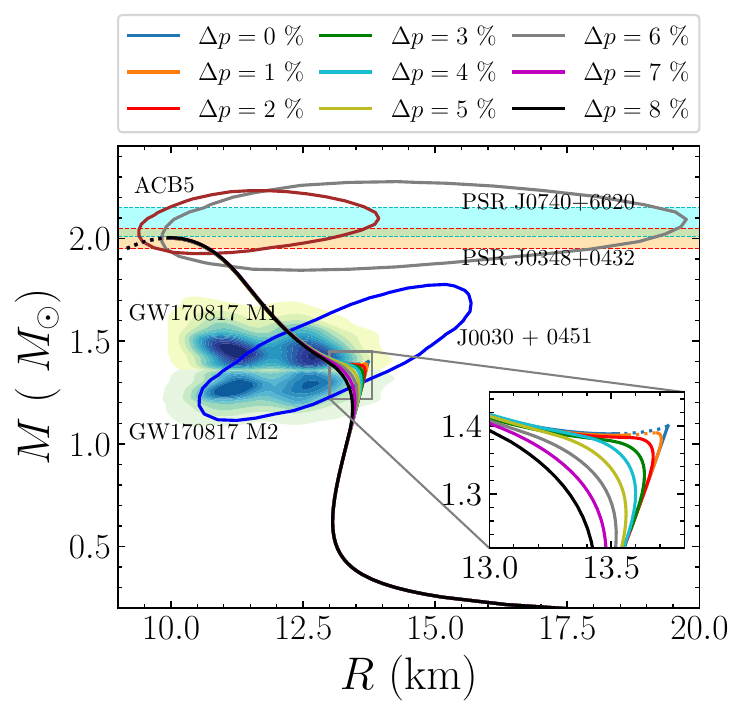}%
  \label{subfig:mr_ACB5}%
}
\caption{Same as ~\cref{fig:EOS_MR} but for the ACB5 parametrized EOS model.}
\label{fig:EOS_MR_ACB5}
\end{figure*}

\begin{figure}
    \centering
    \includegraphics[width=\linewidth]{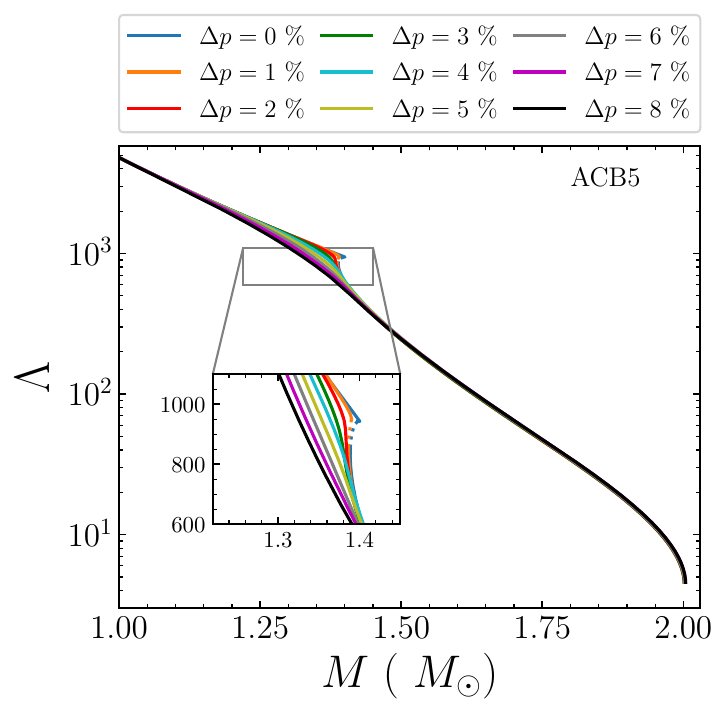}
    \caption{$M-\Lambda$  relations corresponding to the EOSs displayed in ~\cref{subfig:EOS_ACB5}.}
    \label{subfig:mlam_ACB5}
\end{figure}

Depending upon the onset of phase transition, an early phase transition from hadron to quark matter can result in the existence of low-mass twin stars. In our previous discussions, particularly in~\cref{sec:asteroseismology,sec:detection}, we explored various perspectives on the detectability of high-mass twin stars. Before we conclude, we also examine our methodology in the context of low-mass twin stars using the ACB5 parameterized EOS model (see detailed  description in ~\citep{Paschalidis2018}). We present the Equation of State (EOS) and the corresponding Mass-Radius ($M-R$) relations for the ACB5 model in ~\cref{subfig:EOS_ACB5,subfig:mr_ACB5} respectively. Notably, compared to the ACB4 EOS model, the ACB5 EOS model indicates a phase transition occurring at a lower density, with the transition onset at around 1.4$M_{\odot}$. For the ACB5 EOS model, the second and third families of compact stars merge into a single branch for $\Delta p \geq 2\%$. Furthermore, we illustrate the variation of tidal deformability with mass for the ACB5 model in ~\cref{subfig:mlam_ACB5}. Interestingly, the differences in compact star radii ($\Delta R$) and tidal deformability ($\Delta \Lambda$) between hadronic neutron stars (NS) and their twin companions are significantly reduced compared to the high-mass twins.
In an optimistic scenario involving the detection of binaries with next-generation gravitational wave (GW) detectors, it becomes possible to measure $\Delta \Lambda$ within approximately 15\% ~\citep{Landry2022}. This indicates that hybrid stars can be distinguished using these advanced GW detectors, where the tidal deformability $\Lambda$ differs by more than 15\% compared to their hadronic companions.

\begin{figure*}
\centering 

\subfloat[]{%
  \includegraphics[width=0.5\textwidth]{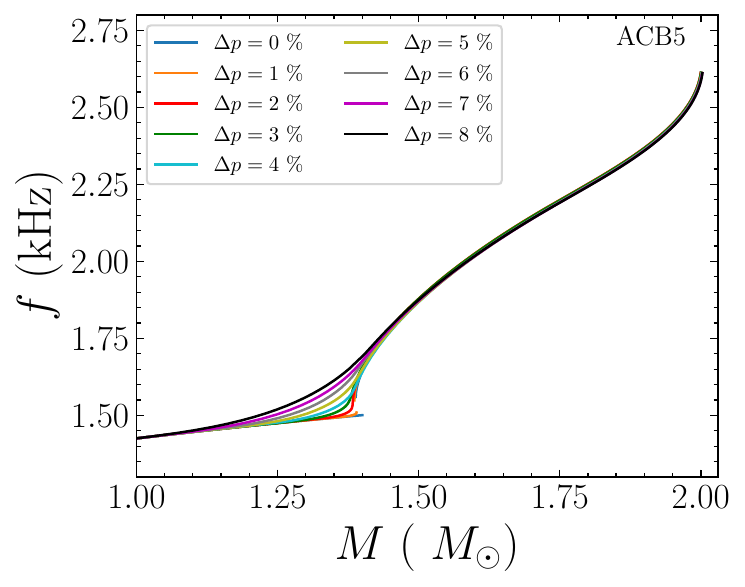}%
  \label{subfig:f_m_ACB5}%
}
\subfloat[]{%
  \includegraphics[width=0.5\textwidth]{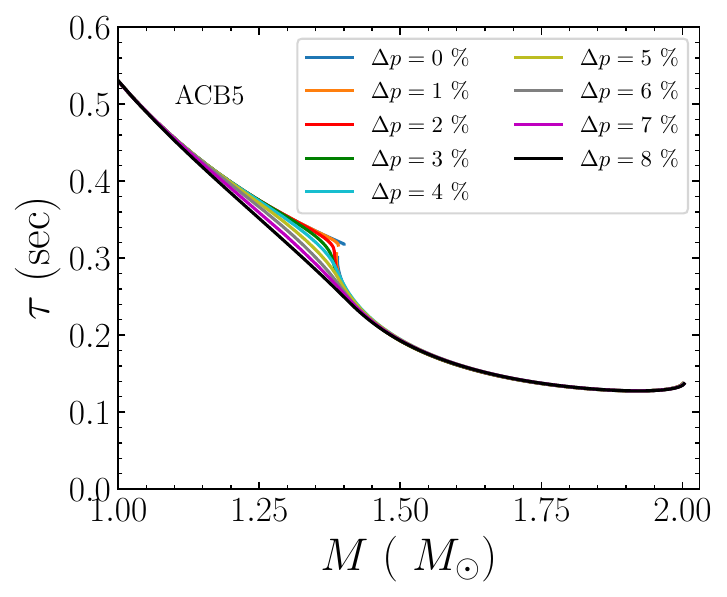}%
  \label{subfig:m_tau_ACB5}%
}\qquad
\caption{Same as ~\cref{fig:ftau_m} but for the ACB5 EOS model displayed in ~\cref{subfig:EOS_ACB5}.} 
\label{fig:ftau_m_ACB5}
\end{figure*}

We display the $f$-mode characteristics for the ACB5 Model in ~\cref{fig:ftau_m_ACB5}. To distinguish the low-mass twins, one needs a precise measurement of mode parameters simultaneously with the mass measurement. Though GW from the binary system simultaneously measures mass and mode frequency, the more significant errors in the $f$-mode frequency~\cite{Williams2022} make the distinguishability of low-mass twins more challenging. For the ACB5 EOS model with $\Delta p=0\%$, the $f$-mode frequency of the third family twin star differs by 5\%  compared to its second family pair at the onset $M\sim 1.4 M_{\odot}$. Hence, a future detection of the $f$-mode frequency of the pulsars with precisely known mass (from other observations, such as radio) within $\leq 5\%$ may put insights into the existence of low-mass twins. Furthermore, from detecting $f$-mode GWs from the pulsars (nearby pulsars are more likely) with unknown macroscopic properties, the conclusion can be made following~\cref{sec:detection}: the detection of $f$-mode GWs with ET can measure the compact star radius $R$ up to $\sim 2\%$ (at a 90\% CI, i.e., $\sigma_{R,90}$). To completely distinguish the twins, the radii of the twins should be separated at least by $2\sigma_{R,90}$ such that the two posterior distributions corresponding to the measurement of the radii of twins do not overlap with each other \footnote{ Though, one might need to perform a proper Bayesian methodology or the statistic tests such as $K-S$ test to test that the distributions are different, several conclusions can be made looking at the posterior distributions or their overlaps.}. Twins  separated by radius $\Delta R$  greater than $2\sigma_{R,90}$ can be distinguishable. Hence, from the detection of $f$-mode GWs from two compact objects having the same mass with   $\frac{2\sigma_{R,90}}{\Delta R} \geq 1$, one might clearly distinguish the twins. Looking at ~\cref{fig:f_tau_and_mr_corner} of ~\cref{sec:detection} we have $2\sigma_{R,90}\sim 0.4$ km for the $f$-mode detection with ET, which concludes that twins having separation $\geq 0.4$km can be distinguished from the future $f$-mode observations.

The investigations of ~\cite{David2021} demonstrated the possibilities of the existence of twin stars at different onset masses, and the results can be used to comment on the detection of twins at other mass regions using $f$-mode GW observations. Our analysis suggests that confirming the presence of low-mass twins and identifying the earliest phase transition may be less feasible even with $f$-mode GW detection compared to high-mass twins. However, various population studies have indicated that a compact star mass distribution peak occurs around $\sim 1.4 M_{\odot}$~\citep{Lucas2023} and given the statistically significant number of observations near this mass range, combining data from different observations can offer valuable insights into the existence of low-mass twin stars (if they indeed exist).

\subsection{ Twin Stars in Binary System}\label{sec:twinstar_binary}

The binary neutron star system provides an excellent scenario for measuring NS mass and frequency. However, $f$-mode parameters can be constrained for GW events in binary systems having high SNR by the next generation GW detectors~\citep{Williams2022}, while  the leading order parameters are chirp mass $\mathcal{M}_c$ (see ~\cref{eqn:chirpmass}) and tidal parameter $\tilde{\Lambda}$~(see ~\cref{eqn:ltilde}) can be well constrained for events detected by both current and next generation GW detectors. As the measurement of  $\mathcal{M}_c$ and  $\tilde{\Lambda}$ can be used to comment on the presence of strong or crossover phase transition, additional information  regarding the nature of the phase transition can be addressed by the simultaneous measurement of mass and  $f$-mode frequency. Considering a series of detections, ~\cite{Landry2022} recently discussed the prospects of the detections of twin stars using  the jump in the $M-\Lambda$ plane, and the  $\tilde{\Lambda}-\mathcal{M}_c$ behavior  has been used to comment on the presence of the twin branch, where
\begin{align}
    \mathcal{M}_c&=\frac{(m_1 m_2)^{3/5}}{(m_1+m_2)^{1/5}} \label{eqn:chirpmass}& \\
    \tilde{\Lambda}&=\frac{16}{13}\l[ \frac{(m_1+12m_2)m_1^4\Lambda_1}{(m_1+m_2)^{1/5}} + 1\xleftarrow{}\xrightarrow{} 2 \r] &  \label{eqn:ltilde} \\
    C_{DT} &= -\frac{1}{X_1X_2}\l[\frac{\Lambda_1}{(m_1\omega_1)^2} X_1^6(155-147X_1) + 1\xleftarrow{}\xrightarrow{} 2 \r] & \label{eqn:cdt}
\end{align}~.
In \cref{eqn:chirpmass,eqn:ltilde,eqn:cdt}, $m_i,\Lambda_i,\omega_i$ are the mass, quadruple tidal deformability and $f$-mode angular frequency of the $i^{th}$ companion, respectively.  $X_i=m_i/(m_1+m_2)$.

\begin{figure*}
    \subfloat[]{%
  \centering
  \includegraphics[width=0.42\linewidth]{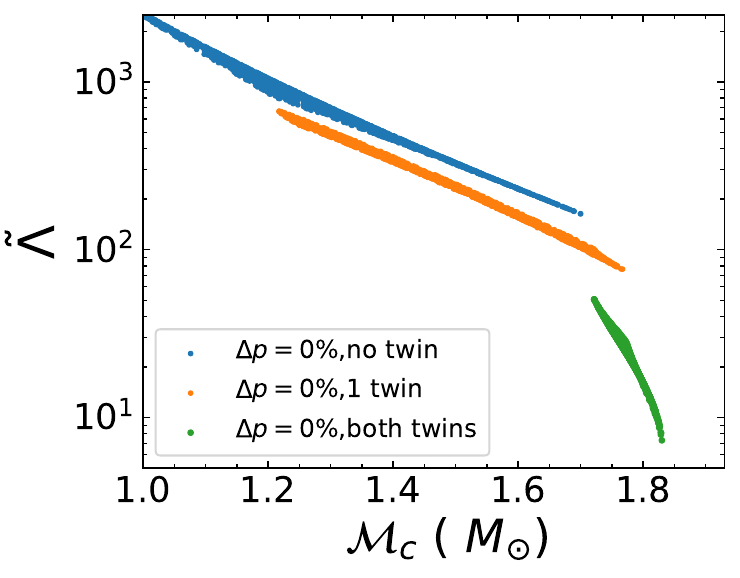}
    \label{fig:mc_ltildedp0}
    }
    \subfloat[]{
      \includegraphics[width=0.42\linewidth]{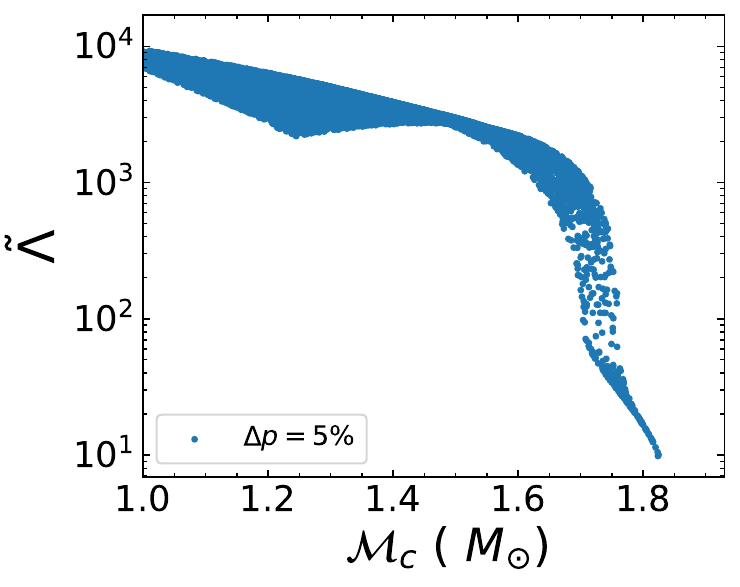}  
  \label{fig:mc_ltildedp5}
    }
    \caption{(a) The binary parameter $\tilde{\Lambda}$ as a function of the binary chirp mass $\mathcal{M}_c$ for the EOS $\Delta p=0\%$. The label `no twin' represents the case when none of the stars in the binary contains a twin star or both are sampled  to the second family. Points labeled as `1 twin' contains one star from the second family and another from the third family. Finally, the scatter points with the label `both twins' contains both binaries with stars from the third family only. (b) Same as (a) but the hybrid  EOS with $\Delta p=5\%$ is considered here. }
    \label{fig:binary_mltilde}
    \end{figure*}

\begin{figure*}
    \subfloat[]{%
  \centering
  \includegraphics[width=0.42\linewidth]{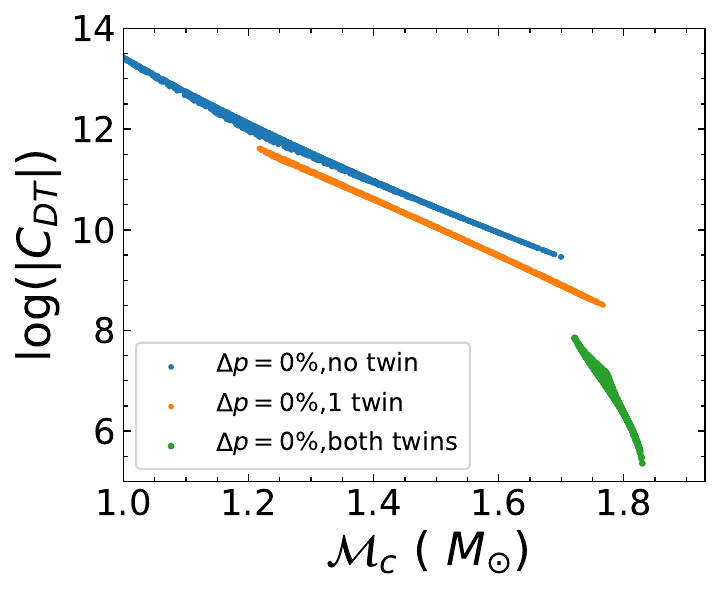}  
  \label{fig:mc_cdtdp0}
    }
    \subfloat[]{
      \includegraphics[width=0.42\linewidth]{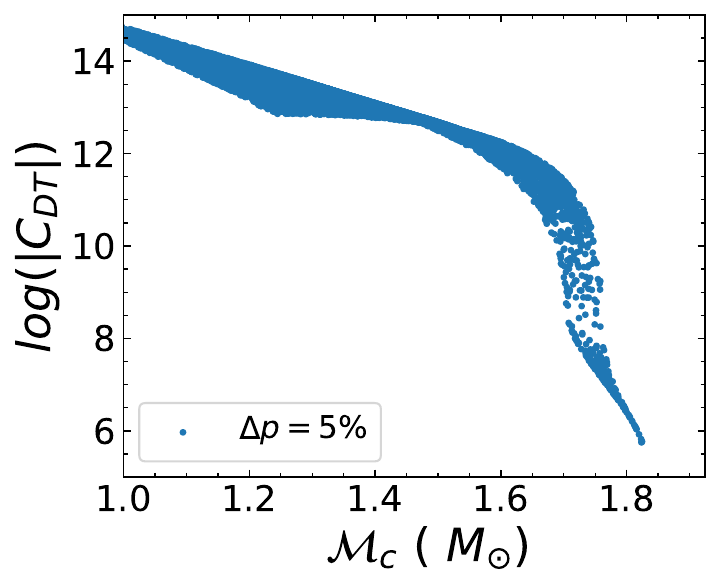} 
  \label{fig:mc_cdtdp5}
    }
    \caption{ Same as figure \cref{fig:binary_mltilde} but for the dynamical tidal parameter $C_{DT}$. }
    \label{fig:binary_mcdt}
    \end{figure*}

The $f$-mode frequency may be detected within a few 100 Hz  from a BNS event  in the era of  XG  GW detectors ~\cite{Williams2022} through the dynamical tidal parameter $C_{DT}$ (see ~\cref{eqn:cdt}). Simultaneously, the leading order tidal parameters would be more precisely measured. Looking at ~\cref{fig:mc_ltildedp0}, it is clear that there is a jump in $\tilde{\Lambda}-\mathcal{M}_c$ plane for binaries with at least one twin star, compared to the binaries with only stars from the second family branch. Similar behavior can be seen in ~\cref{fig:mc_cdtdp0} for the dynamical parameter $C_{DT}$ involving the $f$-mode frequency. However, if one considers an EOS with a crossover-like phase transition and with no discontinuous twin branch (say the EOS with $\Delta p=5\%$), there is no discontinuous jump in  $\tilde{\Lambda}$ or $C_{DT}$ as seen for the EOS with $\Delta p=0\%$ (see ~\cref{fig:mc_cdtdp5}). So the presence of the jump in $\tilde{\Lambda}-\mathcal{M}_c$ or $C_{DT}-\mathcal{M}_c$ can help us indicate the presence of a twin branch or the presence of a strong phase transition. 

\section{Discussions}
\label{sec:discussions}

In this work, we investigated for the first time whether future detection of gravitational waves from $f$-mode oscillations could allow us to probe the nature of the hadron-quark phase transition in NSs. {The nature of the phase transition is intimately related to the stiffness of dense matter. The effect of the sharp first-order phase transitions it to soften matter just as the appearance of pasta phases in our study does. On the other hand, crossover phase transitions might either soften or stiffen matter at the phase transition, see~\citep{Sotani:2023zkk} for examples of the latter. The general tendency of the $f$-modes curves as a function of stellar mass in the case of stiffer matter is to lower their values with respect to the pure hadronic stars case as opposed to increased values for the case of softer matter, as can be seen in figure 4 of ~\citep{Sotani:2023zkk} for stiffening matter and in the left panel of figure~\ref{fig:ftau_m} of this work for softening matter. In our study} we employed a recently developed phenomenological interpolation scheme from~\cite{Abgaryan2018} to mimic the thermodynamic behaviour of the mixed phase via pasta phases, in which the nature of the phase transition is described in terms of a single parameter, the pressure increment ($\Delta p$) at critical chemical potential. Within this EOS scheme, we calculated the global NS properties such as mass, radius and tidal deformability as well as $f$-mode characteristics in full general relativistic framework. We systematically investigated how these observable NS properties are affected by the variation in parameter $\Delta p$, i.e., in going from a sharp ($\Delta p =0)$ to a smooth ($\Delta p=8\%$) phase transition.  
\\

For the case of isolated NSs, we then used Universal Relations (URs) to recover stellar properties such as mass and radius considering future detection of $f$-modes excited by NS glitches. We further analysed whether, given the uncertainties in the URs, the measurement of masses and radii can allow us to comment on the nature of the hadron-quark phase transition. We concluded that if the $f$-mode characteristics are precisely known, for a given UR one may be able to distinguish between a strong phase transition ($\Delta p=0$) and a smooth one ($\Delta p=8\%$), but the distinguishability depends on the value of $\Delta p$. We further considered uncertainties in the $f$-mode observations and found that this would further decrease the distinguishability between the mixed-phase scenarios. However, in all the cases, twin stars (corresponding to the strong phase transition $\Delta p=0$) can be clearly distinguished from a normal NS. 
\\

We also examined whether GW observations from $f$-modes can be used to comment on the presence or distinguishability of the low mass twin stars. Contrary to high mass twins, we discover that applying the same methodology it becomes more challenging to detect low mass twin stars when the $f$-mode detections are made from compact stars of unknown mass. We go over various astrophysical observational scenarios and estimate the amount of observational precision that would be necessary to distinguish between twins at different twin star onset masses.

For the case of binary NSs, we probed whether future observations of dynamical $f$-modes can lead to constraints on the EOS. We found that although one may be able to distinguish NSs from twins using the detection of the dynamical tidal parameter, the constraints from the tidal parameter in the case of binary NSs provide better evidence for the existence of twin stars, which supports the scenario of a sharp hadron-quark phase transition. 
\\

Previous studies~\cite{AyriyanEPJ2018,Pereira_2022,Landry2022} attempted to differentiate between the strong hadron-quark phase transition or smooth crossover by investigating their effects on the mass, radius or tidal deformability. \citep{Pereira_2022} concluded that even considering the most optimistic case for future generation GW detectors, distinguishing a sharp phase transition from a mixed state may be observationally challenging. Landry and Chakravarti~\cite{Landry2022} analyzed the number of binary NS observations required to infer the existence of twin stars from the measurement of tidal deformation.  Recently, in the work of ~\cite{Suleiman2024}, the broadening of a few relevant  URs due to the consideration of different physical constraints and the impact on future NS measurements is discussed. The effect of twin stars or hybrid stars on the URs involving the moment of inertia or relevant for GW analysis from binary systems is subject to future investigation.
\\

It is of great interest to the GW community to calculate the quasi-normal modes of hybrid stars relevant for third-generation detectors~\citep{Hild_2011,CE} or the planned GW mission NEMO~\citep{nemo}. Interestingly,~\cite{Sotani:2023zkk} find that the fundamental $f$-mode frequencies with Quark-hadron crossover EOS basically are smaller and the 1st pressure p1-mode frequencies with
QHC EOS are larger than those with hadronic EOS and, moreover, are able to distinguish between these two possibilities using the so-called universal relations.
It was suggested that oscillation modes in NSs could provide smoking-gun evidence for the nature of the mixed phase. A few recent studies also attempted to study the effects of the mixed phase on NS oscillation modes such as g-modes~\cite{Constantinou2023} and w-modes~\citep{Sandoval2022,Sandoval2022b}, however, they did not comment on their detectability from GW observations.  The results of our investigation demonstrate that for future $f$-mode detections from isolated NSs, whether or not one may be able to differentiate between sharp and smooth phase transitions depends on the sharpness of the phase transition as well as uncertainties in the observations and the universal relations. Improved universal relations and high-precision measurements may provide hope to be able to comment conclusively about the existence of a mixed phase in the NS interior. 
 \\


\section*{Acknowledgements}

B.K.P thanks Swarnim Shrike for the useful discussions they had regarding this work. {D.E.A.C. acknowledges fruitful discussions with Alexander Ayriyan and Noshad Khosravi Largani regarding the Bayesian approach and $f$-mode analysis, respectively. Moreover, he also acknowledges support from the program Excellence Initiative–Research University of the University of Wroclaw of the Ministry of Education and Science.}

\section*{Data Availability}
Data can be available upon request.



\bibliographystyle{mnras}
\bibliography{Pradhan} 

\begin{thebibliography}{}
\makeatletter
\relax
\def\mn@urlcharsother{\let\do\@makeother \do\$\do\&\do\#\do\^\do\_\do\%\do\~}
\def\mn@doi{\begingroup\mn@urlcharsother \@ifnextchar [ {\mn@doi@}
  {\mn@doi@[]}}
\def\mn@doi@[#1]#2{\def\@tempa{#1}\ifx\@tempa\@empty \href
  {http://dx.doi.org/#2} {doi:#2}\else \href {http://dx.doi.org/#2} {#1}\fi
  \endgroup}
\def\mn@eprint#1#2{\mn@eprint@#1:#2::\@nil}
\def\mn@eprint@arXiv#1{\href {http://arxiv.org/abs/#1} {{\tt arXiv:#1}}}
\def\mn@eprint@dblp#1{\href {http://dblp.uni-trier.de/rec/bibtex/#1.xml}
  {dblp:#1}}
\def\mn@eprint@#1:#2:#3:#4\@nil{\def\@tempa {#1}\def\@tempb {#2}\def\@tempc
  {#3}\ifx \@tempc \@empty \let \@tempc \@tempb \let \@tempb \@tempa \fi \ifx
  \@tempb \@empty \def\@tempb {arXiv}\fi \@ifundefined
  {mn@eprint@\@tempb}{\@tempb:\@tempc}{\expandafter \expandafter \csname
  mn@eprint@\@tempb\endcsname \expandafter{\@tempc}}}

\bibitem[\protect\citeauthoryear{Abbott et~al.,}{Abbott
  et~al.}{2017a}]{AbbottPRL119}
Abbott B.,  et~al., 2017a, \mn@doi [Physical Review Letters]
  {10.1103/physrevlett.119.161101}, 119

\bibitem[\protect\citeauthoryear{Abbott et~al.,}{Abbott
  et~al.}{2017b}]{AbbottAJL848}
Abbott B.~P.,  et~al., 2017b, \mn@doi [The Astrophysical Journal]
  {10.3847/2041-8213/aa91c9}, 848, L12

\bibitem[\protect\citeauthoryear{Abbott et~al.,}{Abbott
  et~al.}{2018}]{AbbottPRL121}
Abbott B.,  et~al., 2018, \mn@doi [Physical Review Letters]
  {10.1103/physrevlett.121.161101}, 121

\bibitem[\protect\citeauthoryear{Abbott et~al.,}{Abbott
  et~al.}{2019a}]{AbbottPRX}
Abbott B.,  et~al., 2019a, \mn@doi [Physical Review X]
  {10.1103/physrevx.9.011001}, 9

\bibitem[\protect\citeauthoryear{Abbott et~al.,}{Abbott
  et~al.}{2019b}]{Abbott_2019}
Abbott B.~P.,  et~al., 2019b, \mn@doi [The Astrophysical Journal]
  {10.3847/1538-4357/ab0e15}, 874, 163

\bibitem[\protect\citeauthoryear{Abbott, Abbott  \& Abbott~et al}{Abbott
  et~al.}{2020}]{Abbott2020}
Abbott B.~P.,  Abbott R.,   Abbott~et al T.~D.,  2020, \mn@doi [Living Reviews
  in Relativity] {10.1007/s41114-020-00026-9}, 23, 3

\bibitem[\protect\citeauthoryear{{Abbott} et~al.,}{{Abbott}
  et~al.}{2021}]{AbbottPRD104}
{Abbott} R.,  et~al., 2021, \mn@doi [\prd] {10.1103/PhysRevD.104.122004}, \href
  {https://ui.adsabs.harvard.edu/abs/2021PhRvD.104l2004A} {104, 122004}

\bibitem[\protect\citeauthoryear{{Abbott} et~al.,}{{Abbott}
  et~al.}{2022a}]{AbbottFRB2022}
{Abbott} R.,  et~al., 2022a, \mn@doi [arXiv e-prints]
  {10.48550/arXiv.2203.12038}, \href
  {https://ui.adsabs.harvard.edu/abs/2022arXiv220312038T} {p. arXiv:2203.12038}

\bibitem[\protect\citeauthoryear{{Abbott} et~al.,}{{Abbott}
  et~al.}{2022b}]{AbbottLVK2022}
{Abbott} R.,  et~al., 2022b, \mn@doi [arXiv e-prints]
  {10.48550/arXiv.2210.10931}, \href
  {https://ui.adsabs.harvard.edu/abs/2022arXiv221010931T} {p. arXiv:2210.10931}

\bibitem[\protect\citeauthoryear{{Abgaryan}, {Alvarez-Castillo}, {Ayriyan},
  {Blaschke}  \& {Grigorian}}{{Abgaryan} et~al.}{2018}]{Abgaryan2018}
{Abgaryan} V.,  {Alvarez-Castillo} D.,  {Ayriyan} A.,  {Blaschke} D.,
  {Grigorian} H.,  2018, \mn@doi [Universe] {10.3390/universe4090094}, \href
  {https://ui.adsabs.harvard.edu/abs/2018Univ....4...94A} {4, 94}

\bibitem[\protect\citeauthoryear{Ackley et~al.,}{Ackley et~al.}{2020}]{nemo}
Ackley K.,  et~al., 2020, \mn@doi [Publications of the Astronomical Society of
  Australia] {10.1017/pasa.2020.39}, 37, e047

\bibitem[\protect\citeauthoryear{{Alvarez-Castillo}}{{Alvarez-Castillo}}{2021}]{David2021}
{Alvarez-Castillo} D.~E.,  2021, \mn@doi [Astronomische Nachrichten]
  {10.1002/asna.202113910}, \href
  {https://ui.adsabs.harvard.edu/abs/2021AN....342..234A} {342, 234}

\bibitem[\protect\citeauthoryear{Alvarez-Castillo \& Blaschke}{Alvarez-Castillo
  \& Blaschke}{2017}]{David2017}
Alvarez-Castillo D.~E.,  Blaschke D.~B.,  2017, \mn@doi [Phys. Rev. C]
  {10.1103/PhysRevC.96.045809}, 96, 045809

\bibitem[\protect\citeauthoryear{{Alvarez-Castillo}, {Blaschke}  \&
  {Typel}}{{Alvarez-Castillo} et~al.}{2017}]{David2017b}
{Alvarez-Castillo} D.,  {Blaschke} D.,   {Typel} S.,  2017, \mn@doi
  [Astronomische Nachrichten] {10.1002/asna.201713433}, \href
  {https://ui.adsabs.harvard.edu/abs/2017AN....338.1048A} {338, 1048}

\bibitem[\protect\citeauthoryear{Andersson}{Andersson}{2021}]{Andersson2021}
Andersson N.,  2021, \mn@doi [Universe] {10.3390/universe7040097}, 7

\bibitem[\protect\citeauthoryear{Andersson \& Comer}{Andersson \&
  Comer}{2001}]{Andersson2001}
Andersson N.,  Comer G.~L.,  2001, \mn@doi [Phys. Rev. Lett.]
  {10.1103/PhysRevLett.87.241101}, 87, 241101

\bibitem[\protect\citeauthoryear{Andersson \& Ho}{Andersson \&
  Ho}{2018}]{Andersson2018}
Andersson N.,  Ho W. C.~G.,  2018, \mn@doi [Phys. Rev. D]
  {10.1103/PhysRevD.97.023016}, 97, 023016

\bibitem[\protect\citeauthoryear{Andersson \& Kokkotas}{Andersson \&
  Kokkotas}{1996}]{Andersson96}
Andersson N.,  Kokkotas K.~D.,  1996, \mn@doi [Phys. Rev. Lett.]
  {10.1103/PhysRevLett.77.4134}, 77, 4134

\bibitem[\protect\citeauthoryear{Andersson \& Kokkotas}{Andersson \&
  Kokkotas}{1998}]{Andersson98}
Andersson N.,  Kokkotas K.~D.,  1998, \mn@doi [Mon. Not. Roy. Astron. Soc.]
  {10.1046/j.1365-8711.1998.01840.x}, 299, 1059

\bibitem[\protect\citeauthoryear{{Ashton} et~al.,}{{Ashton}
  et~al.}{2019}]{Bilby}
{Ashton} G.,  et~al., 2019, \mn@doi [The Astrophysical Journal Supplement
  Series] {10.3847/1538-4365/ab06fc}, \href
  {https://ui.adsabs.harvard.edu/abs/2019ApJS..241...27A} {241, 27}

\bibitem[\protect\citeauthoryear{{Ayriyan} \& {Grigorian}}{{Ayriyan} \&
  {Grigorian}}{2018}]{AyriyanEPJ2018}
{Ayriyan} A.,  {Grigorian} H.,  2018, in European Physical Journal Web of
  Conferences. p. 03003 (\mn@eprint {arXiv} {1710.05637}),
  \mn@doi{10.1051/epjconf/201817303003}

\bibitem[\protect\citeauthoryear{Ayriyan, Bastian, Blaschke, Grigorian, Maslov
  \& Voskresensky}{Ayriyan et~al.}{2018}]{Ayriyan}
Ayriyan A.,  Bastian N.-U.,  Blaschke D.,  Grigorian H.,  Maslov K.,
  Voskresensky D.~N.,  2018, \mn@doi [Phys. Rev. C]
  {10.1103/PhysRevC.97.045802}, 97, 045802

\bibitem[\protect\citeauthoryear{{Ayriyan}, {Blaschke}, {Grunfeld},
  {Alvarez-Castillo}, {Grigorian}  \& {Abgaryan}}{{Ayriyan}
  et~al.}{2021}]{Ayriyan2021b}
{Ayriyan} A.,  {Blaschke} D.,  {Grunfeld} A.~G.,  {Alvarez-Castillo} D.,
  {Grigorian} H.,   {Abgaryan} V.,  2021, \mn@doi [European Physical Journal A]
  {10.1140/epja/s10050-021-00619-0}, \href
  {https://ui.adsabs.harvard.edu/abs/2021EPJA...57..318A} {57, 318}

\bibitem[\protect\citeauthoryear{Bauswein \& Janka}{Bauswein \&
  Janka}{2012}]{Bauswein2012}
Bauswein A.,  Janka H.-T.,  2012, \mn@doi [Phys. Rev. Lett.]
  {10.1103/PhysRevLett.108.011101}, 108, 011101

\bibitem[\protect\citeauthoryear{Bauswein, Bastian, Blaschke, Chatziioannou,
  Clark, Fischer  \& Oertel}{Bauswein et~al.}{2019}]{Bauswein:2018bma}
Bauswein A.,  Bastian N.-U.~F.,  Blaschke D.~B.,  Chatziioannou K.,  Clark
  J.~A.,  Fischer T.,   Oertel M.,  2019, \mn@doi [Phys. Rev. Lett.]
  {10.1103/PhysRevLett.122.061102}, 122, 061102

\bibitem[\protect\citeauthoryear{Baym, Hatsuda, Kojo, Powell, Song  \&
  Takatsuka}{Baym et~al.}{2018}]{Baym:2017whm}
Baym G.,  Hatsuda T.,  Kojo T.,  Powell P.~D.,  Song Y.,   Takatsuka T.,  2018,
  \mn@doi [Rept. Prog. Phys.] {10.1088/1361-6633/aaae14}, 81, 056902

\bibitem[\protect\citeauthoryear{Blaschke \& Alvarez-Castillo}{Blaschke \&
  Alvarez-Castillo}{2020}]{Blaschke:2018pva}
Blaschke D.,  Alvarez-Castillo D.,  2020, \mn@doi [Eur. Phys. J. A]
  {10.1140/epja/s10050-020-00111-1}, 56, 124

\bibitem[\protect\citeauthoryear{Blaschke, Alvarez~Castillo, Ayriyan,
  Grigorian, Largani  \& Weber}{Blaschke et~al.}{2020a}]{Blaschke2020}
Blaschke D.,  Alvarez~Castillo D.~E.,  Ayriyan A.,  Grigorian H.,  Largani
  N.~K.,   Weber F.,  2020a, Astrophysical Aspects of General Relativistic Mass
  Twin Stars.
pp 207--256, \mn@doi{10.1142/9789813277342_0007}

\bibitem[\protect\citeauthoryear{Blaschke, Ayriyan, Alvarez-Castillo  \&
  Grigorian}{Blaschke et~al.}{2020b}]{Blaschke20206}
Blaschke D.,  Ayriyan A.,  Alvarez-Castillo D.~E.,   Grigorian H.,  2020b,
  \mn@doi [Universe] {10.3390/universe6060081}, 6, 81

\bibitem[\protect\citeauthoryear{Chan, Sham, Leung  \& Lin}{Chan
  et~al.}{2014}]{Chan2014}
Chan T.~K.,  Sham Y.-H.,  Leung P.~T.,   Lin L.-M.,  2014, \mn@doi [Phys. Rev.
  D] {10.1103/PhysRevD.90.124023}, 90, 124023

\bibitem[\protect\citeauthoryear{Chandrasekhar \& Ferrari}{Chandrasekhar \&
  Ferrari}{1991}]{Chandrasekhar:1991}
Chandrasekhar S.,  Ferrari V.,  1991, \mn@doi [Proc. Roy. Soc. Lond. A]
  {10.1098/rspa.1991.0016}, 432, 247

\bibitem[\protect\citeauthoryear{Chirenti, de Souza  \& Kastaun}{Chirenti
  et~al.}{2015}]{Chirenti2015}
Chirenti C.,  de Souza G.~H.,   Kastaun W.,  2015, \mn@doi [Phys. Rev. D]
  {10.1103/PhysRevD.91.044034}, 91, 044034

\bibitem[\protect\citeauthoryear{Christian}{Christian}{2023}]{Christian2023}
Christian J.-E.,  2023, doctoralthesis, Universit{\"a}tsbibliothek Johann
  Christian Senckenberg, \mn@doi{10.21248/gups.74239}

\bibitem[\protect\citeauthoryear{Christian \& Schaffner-Bielich}{Christian \&
  Schaffner-Bielich}{2020}]{Christian_2020}
Christian J.-E.,  Schaffner-Bielich J.,  2020, \mn@doi [The Astrophysical
  Journal Letters] {10.3847/2041-8213/ab8af4}, 894, L8

\bibitem[\protect\citeauthoryear{Christian \& Schaffner-Bielich}{Christian \&
  Schaffner-Bielich}{2022}]{Christian2022}
Christian J.-E.,  Schaffner-Bielich J.,  2022, \mn@doi [The Astrophysical
  Journal] {10.3847/1538-4357/ac75cf}, 935, 122

\bibitem[\protect\citeauthoryear{Constantinou, Han, Jaikumar  \&
  Prakash}{Constantinou et~al.}{2021}]{Constantinou2021}
Constantinou C.,  Han S.,  Jaikumar P.,   Prakash M.,  2021, \mn@doi [Phys.
  Rev. D] {10.1103/PhysRevD.104.123032}, 104, 123032

\bibitem[\protect\citeauthoryear{Constantinou, Zhao, Han  \&
  Prakash}{Constantinou et~al.}{2023}]{Constantinou2023}
Constantinou C.,  Zhao T.,  Han S.,   Prakash M.,  2023, \mn@doi [Phys. Rev. D]
  {10.1103/PhysRevD.107.074013}, 107, 074013

\bibitem[\protect\citeauthoryear{{Detweiler} \& {Lindblom}}{{Detweiler} \&
  {Lindblom}}{1985}]{Detweiler85}
{Detweiler} S.,  {Lindblom} L.,  1985, \mn@doi [\apj] {10.1086/163127}, \href
  {https://ui.adsabs.harvard.edu/abs/1985ApJ...292...12D} {292, 12}

\bibitem[\protect\citeauthoryear{Espino \& Paschalidis}{Espino \&
  Paschalidis}{2022}]{Espino2022}
Espino P.~L.,  Paschalidis V.,  2022, \mn@doi [Phys. Rev. D]
  {10.1103/PhysRevD.105.043014}, 105, 043014

\bibitem[\protect\citeauthoryear{{Espinoza}, {Lyne}, {Stappers}  \&
  {Kramer}}{{Espinoza} et~al.}{2011}]{jbglitch11}
{Espinoza} C.~M.,  {Lyne} A.~G.,  {Stappers} B.~W.,   {Kramer} M.,  2011,
  \mn@doi [\mnras] {10.1111/j.1365-2966.2011.18503.x}, \href
  {https://ui.adsabs.harvard.edu/abs/2011MNRAS.414.1679E} {414, 1679}

\bibitem[\protect\citeauthoryear{Ferrari, Miniutti  \& Pons}{Ferrari
  et~al.}{2003}]{Ferrari2003}
Ferrari V.,  Miniutti G.,   Pons J.~A.,  2003, \mn@doi [Monthly Notices of the
  Royal Astronomical Society] {10.1046/j.1365-8711.2003.06580.x}, 342, 629

\bibitem[\protect\citeauthoryear{Flores \& Lugones}{Flores \&
  Lugones}{2014}]{Flores_2014}
Flores C.~V.,  Lugones G.,  2014, \mn@doi [Classical and Quantum Gravity]
  {10.1088/0264-9381/31/15/155002}, 31, 155002

\bibitem[\protect\citeauthoryear{Glendenning}{Glendenning}{1992}]{Glendenning1992}
Glendenning N.~K.,  1992, \mn@doi [Phys. Rev. D] {10.1103/PhysRevD.46.1274},
  46, 1274

\bibitem[\protect\citeauthoryear{Hall }{Hall }{2022}]{CE}
Hall  E.~D.,  2022, \mn@doi [Galaxies] {10.3390/galaxies10040090}, 10

\bibitem[\protect\citeauthoryear{Hild et~al.,}{Hild et~al.}{2011}]{Hild_2011}
Hild S.,  et~al., 2011, \mn@doi [Classical and Quantum Gravity]
  {10.1088/0264-9381/28/9/094013}, 28, 094013

\bibitem[\protect\citeauthoryear{Hinderer}{Hinderer}{2008}]{Hinderer}
Hinderer T.,  2008, \mn@doi [The Astrophysical Journal] {10.1086/533487}, 677,
  1216–1220

\bibitem[\protect\citeauthoryear{Ho, Jones, Andersson  \& Espinoza}{Ho
  et~al.}{2020}]{Ho2020}
Ho W. C.~G.,  Jones D.~I.,  Andersson N.,   Espinoza C.~M.,  2020, \mn@doi
  [Phys. Rev. D] {10.1103/PhysRevD.101.103009}, 101, 103009

\bibitem[\protect\citeauthoryear{Jaikumar, Semposki, Prakash  \&
  Constantinou}{Jaikumar et~al.}{2021}]{Jaikumar2021}
Jaikumar P.,  Semposki A.,  Prakash M.,   Constantinou C.,  2021, \mn@doi
  [Phys. Rev. D] {10.1103/PhysRevD.103.123009}, 103, 123009

\bibitem[\protect\citeauthoryear{Keer \& Jones}{Keer \& Jones}{2014}]{Keer2014}
Keer L.,  Jones D.~I.,  2014, \mn@doi [Monthly Notices of the Royal
  Astronomical Society] {10.1093/mnras/stu2123}, 446, 865

\bibitem[\protect\citeauthoryear{{Khosravi Largani}, {Fischer}  \&
  {Bastian}}{{Khosravi Largani} et~al.}{2023a}]{Largani:2023oyk}
{Khosravi Largani} N.,  {Fischer} T.,   {Bastian} N. U.~F.,  2023a, \mn@doi
  [arXiv e-prints] {10.48550/arXiv.2304.12316}, \href
  {https://ui.adsabs.harvard.edu/abs/2023arXiv230412316K} {p. arXiv:2304.12316}

\bibitem[\protect\citeauthoryear{{Khosravi Largani}, {Fischer}, {Shibagaki},
  {Cerd{\'a}-Dur{\'a}n}  \& {Torres-Forn{\'e}}}{{Khosravi Largani}
  et~al.}{2023b}]{Largani:2023kjx}
{Khosravi Largani} N.,  {Fischer} T.,  {Shibagaki} S.,  {Cerd{\'a}-Dur{\'a}n}
  P.,   {Torres-Forn{\'e}} A.,  2023b, \mn@doi [arXiv e-prints]
  {10.48550/arXiv.2311.15992}, \href
  {https://ui.adsabs.harvard.edu/abs/2023arXiv231115992K} {p. arXiv:2311.15992}

\bibitem[\protect\citeauthoryear{Kokkotas, Apostolatos  \& Andersson}{Kokkotas
  et~al.}{2001}]{Kokkotas2001}
Kokkotas K.~D.,  Apostolatos T.~A.,   Andersson N.,  2001, \mn@doi [Monthly
  Notices of the Royal Astronomical Society]
  {10.1046/j.1365-8711.2001.03945.x}, 320, 307

\bibitem[\protect\citeauthoryear{{Kumar} et~al.,}{{Kumar}
  et~al.}{2023a}]{Muses2023}
{Kumar} R.,  et~al., 2023a, \mn@doi [arXiv e-prints]
  {10.48550/arXiv.2303.17021}, \href
  {https://ui.adsabs.harvard.edu/abs/2023arXiv230317021K} {p. arXiv:2303.17021}

\bibitem[\protect\citeauthoryear{{Kumar}, {Malik}, {Mishra}  \&
  {Provid{\^e}ncia}}{{Kumar} et~al.}{2023b}]{Kumar_2023a}
{Kumar} D.,  {Malik} T.,  {Mishra} H.,   {Provid{\^e}ncia} C.,  2023b, \mn@doi
  [arXiv e-prints] {10.48550/arXiv.2306.09277}, \href
  {https://ui.adsabs.harvard.edu/abs/2023arXiv230609277K} {p. arXiv:2306.09277}

\bibitem[\protect\citeauthoryear{Kumar, Mishra  \& Malik}{Kumar
  et~al.}{2023c}]{Kumar_2023}
Kumar D.,  Mishra H.,   Malik T.,  2023c, \mn@doi [Journal of Cosmology and
  Astroparticle Physics] {10.1088/1475-7516/2023/02/015}, 2023, 015

\bibitem[\protect\citeauthoryear{{Landry} \& {Chakravarti}}{{Landry} \&
  {Chakravarti}}{2022}]{Landry2022}
{Landry} P.,  {Chakravarti} K.,  2022, \mn@doi [arXiv e-prints]
  {10.48550/arXiv.2212.09733}, \href
  {https://ui.adsabs.harvard.edu/abs/2022arXiv221209733L} {p. arXiv:2212.09733}

\bibitem[\protect\citeauthoryear{Largani, Fischer, Sedrakian, Cierniak,
  Alvarez-Castillo  \& Blaschke}{Largani et~al.}{2022}]{Largani:2021hjo}
Largani N.~K.,  Fischer T.,  Sedrakian A.,  Cierniak M.,  Alvarez-Castillo
  D.~E.,   Blaschke D.~B.,  2022, \mn@doi [Mon. Not. Roy. Astron. Soc.]
  {10.1093/mnras/stac1916}, 515, 3539

\bibitem[\protect\citeauthoryear{Laskos-Patkos \& Moustakidis}{Laskos-Patkos \&
  Moustakidis}{2023}]{Laskos2023}
Laskos-Patkos P.,  Moustakidis C.~C.,  2023, \mn@doi [Phys. Rev. D]
  {10.1103/PhysRevD.107.123023}, 107, 123023

\bibitem[\protect\citeauthoryear{Leins, Nollert  \& Soffel}{Leins
  et~al.}{1993}]{Leins1993}
Leins M.,  Nollert H.~P.,   Soffel M.~H.,  1993, \mn@doi [Phys. Rev. D]
  {10.1103/PhysRevD.48.3467}, 48, 3467

\bibitem[\protect\citeauthoryear{Lin, Li, Xu, Ko  \& Wen}{Lin
  et~al.}{2011}]{Lin2011}
Lin W.,  Li B.-A.,  Xu J.,  Ko C.~M.,   Wen D.~H.,  2011, \mn@doi [Phys. Rev.
  C] {10.1103/PhysRevC.83.045802}, 83, 045802

\bibitem[\protect\citeauthoryear{{Lindblom} \& {Detweiler}}{{Lindblom} \&
  {Detweiler}}{1983}]{Detweiler83}
{Lindblom} L.,  {Detweiler} S.~L.,  1983, \mn@doi [\apj] {10.1086/190884},
  \href {https://ui.adsabs.harvard.edu/abs/1983ApJS...53...73L} {53, 73}

\bibitem[\protect\citeauthoryear{Lioutas \& Stergioulas}{Lioutas \&
  Stergioulas}{2018}]{Lioutas2018}
Lioutas G.,  Stergioulas N.,  2018, \mn@doi [Gen. Rel. Grav.]
  {10.1007/s10714-017-2331-7}, 50, 12

\bibitem[\protect\citeauthoryear{Lopez, Tiwari, Drago, Keitel, Lazzaro  \&
  Prodi}{Lopez et~al.}{2022}]{Lopez2022}
Lopez D.,  Tiwari S.,  Drago M.,  Keitel D.,  Lazzaro C.,   Prodi G.~A.,  2022,
  \mn@doi [Phys. Rev. D] {10.1103/PhysRevD.106.103037}, 106, 103037

\bibitem[\protect\citeauthoryear{{Manchester}, {Hobbs}, {Teoh}  \&
  {Hobbs}}{{Manchester} et~al.}{2005}]{atnf05}
{Manchester} R.~N.,  {Hobbs} G.~B.,  {Teoh} A.,   {Hobbs} M.,  2005, \mn@doi
  [\aj] {10.1086/428488}, \href
  {https://ui.adsabs.harvard.edu/abs/2005AJ....129.1993M} {129, 1993}

\bibitem[\protect\citeauthoryear{Maslov, Yasutake, Blaschke, Ayriyan,
  Grigorian, Maruyama, Tatsumi  \& Voskresensky}{Maslov
  et~al.}{2019}]{Maslov2019}
Maslov K.,  Yasutake N.,  Blaschke D.,  Ayriyan A.,  Grigorian H.,  Maruyama
  T.,  Tatsumi T.,   Voskresensky D.~N.,  2019, \mn@doi [Phys. Rev. C]
  {10.1103/PhysRevC.100.025802}, 100, 025802

\bibitem[\protect\citeauthoryear{Masuda, Hatsuda  \& Takatsuka}{Masuda
  et~al.}{2013}]{Masuda2013}
Masuda K.,  Hatsuda T.,   Takatsuka T.,  2013, \mn@doi [Progress of Theoretical
  and Experimental Physics] {10.1093/ptep/ptt045}, 2013

\bibitem[\protect\citeauthoryear{Miller et~al.,}{Miller
  et~al.}{2021}]{Miller_2021}
Miller M.~C.,  et~al., 2021, \mn@doi [The Astrophysical Journal Letters]
  {10.3847/2041-8213/ac089b}, 918, L28

\bibitem[\protect\citeauthoryear{Mock \& Joss}{Mock \& Joss}{1998}]{Mock_1998}
Mock P.~C.,  Joss P.~C.,  1998, \mn@doi [The Astrophysical Journal]
  {10.1086/305693}, 500, 374

\bibitem[\protect\citeauthoryear{Monta\~na, Tol\'os, Hanauske  \&
  Rezzolla}{Monta\~na et~al.}{2019}]{Gloria2019}
Monta\~na G.,  Tol\'os L.,  Hanauske M.,   Rezzolla L.,  2019, \mn@doi [Phys.
  Rev. D] {10.1103/PhysRevD.99.103009}, 99, 103009

\bibitem[\protect\citeauthoryear{Oppenheimer \& Volkoff}{Oppenheimer \&
  Volkoff}{1939}]{Oppenheimer}
Oppenheimer J.~R.,  Volkoff G.~M.,  1939, \mn@doi [Phys. Rev.]
  {10.1103/PhysRev.55.374}, 55, 374

\bibitem[\protect\citeauthoryear{Paschalidis, Yagi, Alvarez-Castillo, Blaschke
  \& Sedrakian}{Paschalidis et~al.}{2018}]{Paschalidis2018}
Paschalidis V.,  Yagi K.,  Alvarez-Castillo D.,  Blaschke D.~B.,   Sedrakian
  A.,  2018, \mn@doi [Phys. Rev. D] {10.1103/PhysRevD.97.084038}, 97, 084038

\bibitem[\protect\citeauthoryear{Pereira, Flores  \& Lugones}{Pereira
  et~al.}{2018}]{Pereira2018}
Pereira J.~P.,  Flores C.~V.,   Lugones G.,  2018, \mn@doi [The Astrophysical
  Journal] {10.3847/1538-4357/aabfbf}, 860, 12

\bibitem[\protect\citeauthoryear{Pereira, Bejger, Zdunik  \& Haensel}{Pereira
  et~al.}{2022}]{Pereira_2022}
Pereira J.~P.,  Bejger M.,  Zdunik J.~L.,   Haensel P.,  2022, \mn@doi [Phys.
  Rev. D] {10.1103/PhysRevD.105.123015}, 105, 123015

\bibitem[\protect\citeauthoryear{Pradhan \& Chatterjee}{Pradhan \&
  Chatterjee}{2021}]{Pradhan2021}
Pradhan B.~K.,  Chatterjee D.,  2021, \mn@doi [Phys. Rev. C]
  {10.1103/PhysRevC.103.035810}, 103, 035810

\bibitem[\protect\citeauthoryear{Pradhan, Chatterjee, Lanoye  \&
  Jaikumar}{Pradhan et~al.}{2022}]{Pradhan2022}
Pradhan B.~K.,  Chatterjee D.,  Lanoye M.,   Jaikumar P.,  2022, \mn@doi [Phys.
  Rev. C] {10.1103/PhysRevC.106.015805}, 106, 015805

\bibitem[\protect\citeauthoryear{{Pradhan}, {Pathak}  \&
  {Chatterjee}}{{Pradhan} et~al.}{2023a}]{Pradhan2023b}
{Pradhan} B.~K.,  {Pathak} D.,   {Chatterjee} D.,  2023a, \mn@doi [arXiv
  e-prints] {10.48550/arXiv.2306.04626}, \href
  {https://ui.adsabs.harvard.edu/abs/2023arXiv230604626K} {p. arXiv:2306.04626}

\bibitem[\protect\citeauthoryear{Pradhan, Vijaykumar  \& Chatterjee}{Pradhan
  et~al.}{2023b}]{Pradhan2023}
Pradhan B.~K.,  Vijaykumar A.,   Chatterjee D.,  2023b, \mn@doi [Phys. Rev. D]
  {10.1103/PhysRevD.107.023010}, 107, 023010

\bibitem[\protect\citeauthoryear{Ranea-Sandoval, Guilera, Mariani  \&
  Orsaria}{Ranea-Sandoval et~al.}{2018}]{Sandoval_2018}
Ranea-Sandoval I.~F.,  Guilera O.~M.,  Mariani M.,   Orsaria M.~G.,  2018,
  \mn@doi [Journal of Cosmology and Astroparticle Physics]
  {10.1088/1475-7516/2018/12/031}, 2018, 031

\bibitem[\protect\citeauthoryear{Ranea-Sandoval, Guilera, Mariani  \&
  Lugones}{Ranea-Sandoval et~al.}{2022a}]{Sandoval2022b}
Ranea-Sandoval I.~F.,  Guilera O.~M.,  Mariani M.,   Lugones G.,  2022a,
  \mn@doi [Phys. Rev. D] {10.1103/PhysRevD.106.043025}, 106, 043025

\bibitem[\protect\citeauthoryear{Ranea-Sandoval, Mariani, Lugones  \&
  Guilera}{Ranea-Sandoval et~al.}{2022b}]{Sandoval2022}
Ranea-Sandoval I.~F.,  Mariani M.,  Lugones G.,   Guilera O.~M.,  2022b,
  \mn@doi [Monthly Notices of the Royal Astronomical Society]
  {10.1093/mnras/stac3780}, 519, 3194

\bibitem[\protect\citeauthoryear{Ranea-Sandoval, Mariani, Celi, Rodr\'{\i}guez
  \& Tonetto}{Ranea-Sandoval et~al.}{2023}]{Sandoval2023fmode}
Ranea-Sandoval I.~F.,  Mariani M.,  Celi M.~O.,  Rodr\'{\i}guez M.~C.,
  Tonetto L.,  2023, \mn@doi [Phys. Rev. D] {10.1103/PhysRevD.107.123028}, 107,
  123028

\bibitem[\protect\citeauthoryear{Ravenhall, Pethick  \& Wilson}{Ravenhall
  et~al.}{1983}]{Ravenhall}
Ravenhall D.~G.,  Pethick C.~J.,   Wilson J.~R.,  1983, \mn@doi [Phys. Rev.
  Lett.] {10.1103/PhysRevLett.50.2066}, 50, 2066

\bibitem[\protect\citeauthoryear{Riley et~al.,}{Riley et~al.}{2021}]{Riley2021}
Riley T.~E.,  et~al., 2021, \mn@doi [The Astrophysical Journal Letters]
  {10.3847/2041-8213/ac0a81}, 918, L27

\bibitem[\protect\citeauthoryear{Rodríguez, Ranea-Sandoval, Mariani, Malfatti
  \& Guilera}{Rodríguez et~al.}{2021}]{Rodriguez}
Rodríguez M.~C.,  Ranea-Sandoval I.~F.,  Mariani M.,  Malfatti G.,   Guilera
  O.~M.,  2021, \mn@doi [Astronomische Nachrichten]
  {https://doi.org/10.1002/asna.202113924}, 342, 305

\bibitem[\protect\citeauthoryear{{Seidov}}{{Seidov}}{1971}]{Seidov1971}
{Seidov} Z.~F.,  1971, \sovast, \href
  {https://ui.adsabs.harvard.edu/abs/1971SvA....15..347S} {15, 347}

\bibitem[\protect\citeauthoryear{Shibata}{Shibata}{1994}]{Shibata1994}
Shibata M.,  1994, \mn@doi [Progress of Theoretical Physics]
  {10.1143/ptp/91.5.871}, 91, 871

\bibitem[\protect\citeauthoryear{Shirke, Ghosh  \& Chatterjee}{Shirke
  et~al.}{2023}]{Shirke_2023}
Shirke S.,  Ghosh S.,   Chatterjee D.,  2023, \mn@doi [The Astrophysical
  Journal] {10.3847/1538-4357/acac31}, 944, 7

\bibitem[\protect\citeauthoryear{Sotani \& Kojo}{Sotani \&
  Kojo}{2023}]{Sotani:2023zkk}
Sotani H.,  Kojo T.,  2023, \mn@doi [Phys. Rev. D]
  {10.1103/PhysRevD.108.063004}, 108, 063004

\bibitem[\protect\citeauthoryear{Sotani \& Kumar}{Sotani \&
  Kumar}{2021}]{Sotani2021}
Sotani H.,  Kumar B.,  2021, \mn@doi [Phys. Rev. D]
  {10.1103/PhysRevD.104.123002}, 104, 123002

\bibitem[\protect\citeauthoryear{Sotani, Tominaga  \& Maeda}{Sotani
  et~al.}{2001}]{Sotani2001}
Sotani H.,  Tominaga K.,   Maeda K.-i.,  2001, \mn@doi [Phys. Rev. D]
  {10.1103/PhysRevD.65.024010}, 65, 024010

\bibitem[\protect\citeauthoryear{Speagle}{Speagle}{2020}]{dynesty}
Speagle J.~S.,  2020, \mn@doi [Mon. Not. Roy. Astron. Soc.]
  {10.1093/mnras/staa278}, 493, 3132

\bibitem[\protect\citeauthoryear{Steinhoff, Hinderer, Buonanno  \&
  Taracchini}{Steinhoff et~al.}{2016}]{Steinhoff2016}
Steinhoff J.,  Hinderer T.,  Buonanno A.,   Taracchini A.,  2016, \mn@doi
  [Phys. Rev. D] {10.1103/PhysRevD.94.104028}, 94, 104028

\bibitem[\protect\citeauthoryear{Stergioulas, Bauswein, Zagkouris  \&
  Janka}{Stergioulas et~al.}{2011}]{Stergioulas2011}
Stergioulas N.,  Bauswein A.,  Zagkouris K.,   Janka H.-T.,  2011, \mn@doi
  [Monthly Notices of the Royal Astronomical Society]
  {10.1111/j.1365-2966.2011.19493.x}, 418, 427

\bibitem[\protect\citeauthoryear{{Suleiman} \& {Read}}{{Suleiman} \&
  {Read}}{2024}]{Suleiman2024}
{Suleiman} L.,  {Read} J.,  2024, \mn@doi [arXiv e-prints]
  {10.48550/arXiv.2402.01948}, \href
  {https://ui.adsabs.harvard.edu/abs/2024arXiv240201948S} {p. arXiv:2402.01948}

\bibitem[\protect\citeauthoryear{Tak\'atsy, Kov\'acs, Wolf  \&
  Schaffner-Bielich}{Tak\'atsy et~al.}{2023}]{Takatsy2023}
Tak\'atsy J.,  Kov\'acs P.,  Wolf G.,   Schaffner-Bielich J.,  2023, \mn@doi
  [Phys. Rev. D] {10.1103/PhysRevD.108.043002}, 108, 043002

\bibitem[\protect\citeauthoryear{{Thorne} \& {Campolattaro}}{{Thorne} \&
  {Campolattaro}}{1967}]{Thorne}
{Thorne} K.~S.,  {Campolattaro} A.,  1967, \mn@doi [\apj] {10.1086/149288},
  \href {https://ui.adsabs.harvard.edu/abs/1967ApJ...149..591T} {149, 591}

\bibitem[\protect\citeauthoryear{Tolman}{Tolman}{1939}]{Tolman}
Tolman R.~C.,  1939, \mn@doi [Phys. Rev.] {10.1103/PhysRev.55.364}, 55, 364

\bibitem[\protect\citeauthoryear{Tsaloukidis, Koliogiannis, Kanakis-Pegios  \&
  Moustakidis}{Tsaloukidis et~al.}{2023}]{Tsaloukidis2023}
Tsaloukidis L.,  Koliogiannis P.~S.,  Kanakis-Pegios A.,   Moustakidis C.~C.,
  2023, \mn@doi [Phys. Rev. D] {10.1103/PhysRevD.107.023012}, 107, 023012

\bibitem[\protect\citeauthoryear{Tsui \& Leung}{Tsui \& Leung}{2005}]{Tsui2005}
Tsui L.~K.,  Leung P.~T.,  2005, \mn@doi [Monthly Notices of the Royal
  Astronomical Society] {10.1111/j.1365-2966.2005.08710.x}, 357, 1029

\bibitem[\protect\citeauthoryear{V\"olkel \& Kr\"uger}{V\"olkel \&
  Kr\"uger}{2022}]{Volkel2022}
V\"olkel S.~H.,  Kr\"uger C.~J.,  2022, \mn@doi [Phys. Rev. D]
  {10.1103/PhysRevD.105.124071}, 105, 124071

\bibitem[\protect\citeauthoryear{V\"olkel, Kr\"uger  \& Kokkotas}{V\"olkel
  et~al.}{2021}]{Volkel2021}
V\"olkel S.~H.,  Kr\"uger C.~J.,   Kokkotas K.~D.,  2021, \mn@doi [Phys. Rev.
  D] {10.1103/PhysRevD.103.083008}, 103, 083008

\bibitem[\protect\citeauthoryear{Voskresensky, Yasuhira  \&
  Tatsumi}{Voskresensky et~al.}{2003}]{VOSKRESENSKY2003291}
Voskresensky D.,  Yasuhira M.,   Tatsumi T.,  2003, \mn@doi [Nuclear Physics A]
  {https://doi.org/10.1016/S0375-9474(03)01313-7}, 723, 291

\bibitem[\protect\citeauthoryear{Wen, Li, Chen  \& Zhang}{Wen
  et~al.}{2019}]{Wen}
Wen D.-H.,  Li B.-A.,  Chen H.-Y.,   Zhang N.-B.,  2019, \mn@doi [Phys. Rev. C]
  {10.1103/PhysRevC.99.045806}, 99, 045806

\bibitem[\protect\citeauthoryear{Williams, Pratten  \& Schmidt}{Williams
  et~al.}{2022}]{Williams2022}
Williams N.,  Pratten G.,   Schmidt P.,  2022, \mn@doi [Phys. Rev. D]
  {10.1103/PhysRevD.105.123032}, 105, 123032

\bibitem[\protect\citeauthoryear{Yagi \& Yunes}{Yagi \&
  Yunes}{2013}]{Yagi:2013bca}
Yagi K.,  Yunes N.,  2013, \mn@doi [Science] {10.1126/science.1236462}, 341,
  365

\bibitem[\protect\citeauthoryear{Yasutake, \L{}astowiecki,
  Beni\ifmmode~\acute{c}\else \'{c}\fi{}, Blaschke, Maruyama  \&
  Tatsumi}{Yasutake et~al.}{2014}]{Yasutake2014}
Yasutake N.,  \L{}astowiecki R.,  Beni\ifmmode~\acute{c}\else \'{c}\fi{} S.,
  Blaschke D.,  Maruyama T.,   Tatsumi T.,  2014, \mn@doi [Phys. Rev. C]
  {10.1103/PhysRevC.89.065803}, 89, 065803

\bibitem[\protect\citeauthoryear{{Yim} \& {Jones}}{{Yim} \&
  {Jones}}{2023}]{Yim2023}
{Yim} G.,  {Jones} D.~I.,  2023, \mn@doi [\mnras] {10.1093/mnras/stac3405},
  \href {https://ui.adsabs.harvard.edu/abs/2023MNRAS.518.4322Y} {518, 4322}

\bibitem[\protect\citeauthoryear{Yoshida \& Kojima}{Yoshida \&
  Kojima}{1997}]{Yoshida}
Yoshida S.,  Kojima Y.,  1997, \mn@doi [Monthly Notices of the Royal
  Astronomical Society] {10.1093/mnras/289.1.117}, 289, 117

\bibitem[\protect\citeauthoryear{Zacchi, Tolos  \& Schaffner-Bielich}{Zacchi
  et~al.}{2017}]{Zacchi2017}
Zacchi A.,  Tolos L.,   Schaffner-Bielich J.,  2017, \mn@doi [Phys. Rev. D]
  {10.1103/PhysRevD.95.103008}, 95, 103008

\bibitem[\protect\citeauthoryear{Zerilli}{Zerilli}{1970}]{Zerilli}
Zerilli F.~J.,  1970, \mn@doi [Phys. Rev. Lett.] {10.1103/PhysRevLett.24.737},
  24, 737

\bibitem[\protect\citeauthoryear{Zhao \& Lattimer}{Zhao \&
  Lattimer}{2022}]{Zhao2022b}
Zhao T.,  Lattimer J.~M.,  2022, \mn@doi [Phys. Rev. D]
  {10.1103/PhysRevD.106.123002}, 106, 123002

\bibitem[\protect\citeauthoryear{Zhao, Constantinou, Jaikumar  \& Prakash}{Zhao
  et~al.}{2022}]{Zhao2022}
Zhao T.,  Constantinou C.,  Jaikumar P.,   Prakash M.,  2022, \mn@doi [Phys.
  Rev. D] {10.1103/PhysRevD.105.103025}, 105, 103025

\bibitem[\protect\citeauthoryear{de Sá, Bernardo, Bachega, Rocha, Moraes  \&
  Horvath}{de~Sá et~al.}{2023}]{Lucas2023}
de Sá L.~M.,  Bernardo A.,  Bachega R. R.~A.,  Rocha L.~S.,  Moraes P. H.
  R.~S.,   Horvath J.~E.,  2023, \mn@doi [Galaxies] {10.3390/galaxies11010019},
  11, 19

\makeatother
\end{thebibliography}




\appendix

\section{Additional information related to the Universal Relations}\label{app:URs_add}

\begin{figure*}
\centering
    \subfloat[]{%
  \includegraphics[width=0.45\textwidth]{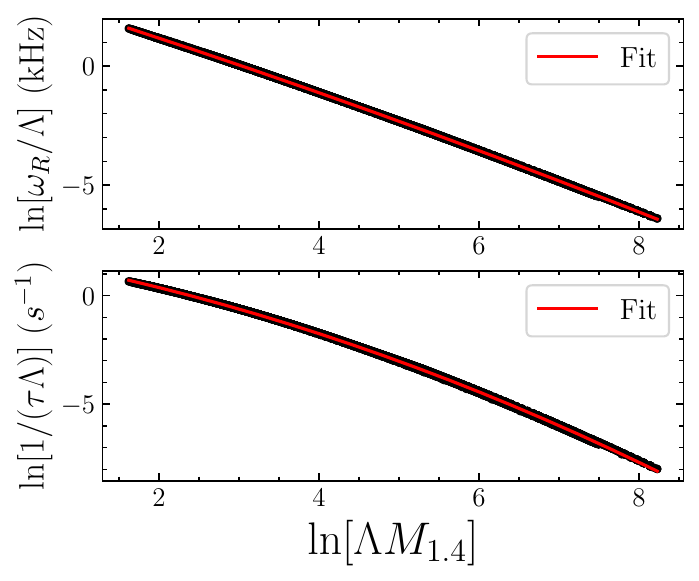}%
  \label{subfig:omega_mlambda}%
}
\subfloat[]{%
  \includegraphics[width=0.45\textwidth]{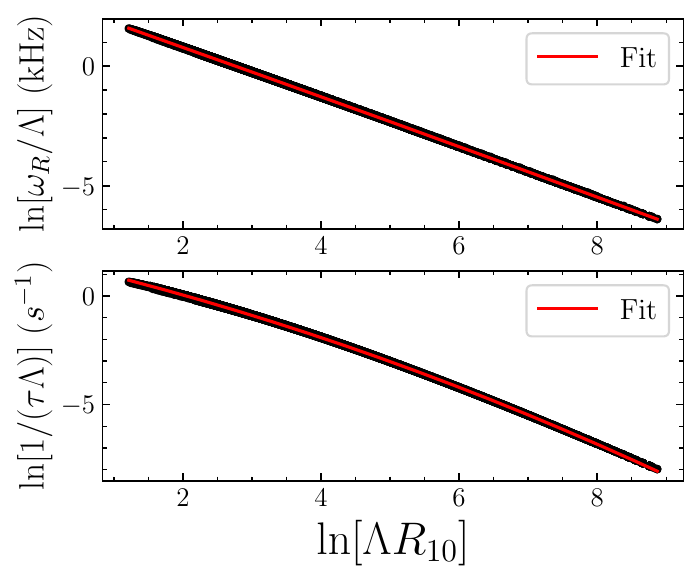}%
  \label{subfig:omega_rlambda}%
}\qquad
    \caption{(\cref{subfig:omega_mlambda}) Universality of  Re($\omega/\Lambda$) (upper panel) and  $1/(\tau \Lambda$) (lower panel) as a function of $\Lambda M_{1.4M_{\odot}}$. The fit relations from ~\cref{eqn:wlambda_mlambda} are also displayed. (\cref{subfig:omega_rlambda}) Same as \cref{subfig:omega_mlambda}, but with
    $\Lambda R_{10}$ as the independent variable and the `Fit' relations correspond to~\cref{eqn:wlambda_radlambda}.}
    \label{fig:tidal_and_fmode}
    \end{figure*}

In Sec.~\ref{sec:URs}, we provided URs for $\l[\frac{\omega_{R,I}}{\Lambda}\r]$ as a function of $\ln\l[M_{1.4M_{\odot}}\Lambda\r]$ in eq.~\ref{eqn:wlambda_mlambda} and $\ln\l[\Lambda R_{10\mathrm{km}}\r]$ in eq.~\ref{eqn:wlambda_radlambda}.
We display the URs for $\l[\frac{\omega_{R,I}}{\Lambda}\r]$ as a function of $\ln\l[M_{1.4M_{\odot}}\Lambda\r]$   in~\cref{subfig:omega_mlambda}. We display the URs among $\omega_{R,I}$ and $\ln\l[\Lambda R_{10\mathrm{km}}\r]$ in ~\cref{subfig:omega_rlambda}. 
    
For comparison with previous works, we also provide the empirical fit relations for frequency  as a function of mean density and scaled damping time as a function of compactness  in~\cref{eqn:f_sqrtden,eqn:tau_comp}, respectively,

\begin{equation}\label{eqn:f_sqrtden}
    f (\rm{kHz})=(39.19\pm0.15) \sqrt{\frac{M}{R^3}}+(0.52\pm 0.006)
\end{equation}
\begin{equation}\label{eqn:tau_comp}
    \frac{R^4}{M^3\tau}=(-0.26\pm5\times 10^{-4}) \frac{M}{R}+(0.082\pm1.1\times 10^{-4})
\end{equation}

Furthermore, there exist URs among mass-scaled angular frequency ($M \omega_R$) and mass-scaled damping time ($M/\tau$) of $f$-modes. These URs can also be used to estimate the NS mass from the measurements of $f,\tau$ from GW events or can be used to constrain $\tau$ from simultaneous measurements of $M,f$ ( possible from a GW event in a binary system). The UR among Re($M\omega)=M\omega_R$ and $Im(M\omega)=M/\tau$ is given in ~\cref{eqn:remomega_immomega},

\begin{align}\label{eqn:remomega_immomega}
\frac{M}{\tau}&=&6.96\times 10^{-6}-4.34\times 10^{-4}(M\omega_R)+  1.48\times 10^{-2} \l(M\omega_R\r)^2 \nonumber \\
& &-8.9\times 10^{-2}\l(M\omega_R\r)^3+0.43\l(M\omega_R\r)^4-1.48\l(M\omega_R\r)^5
\end{align}

\begin{figure}
    \centering
    \includegraphics[width=\linewidth]{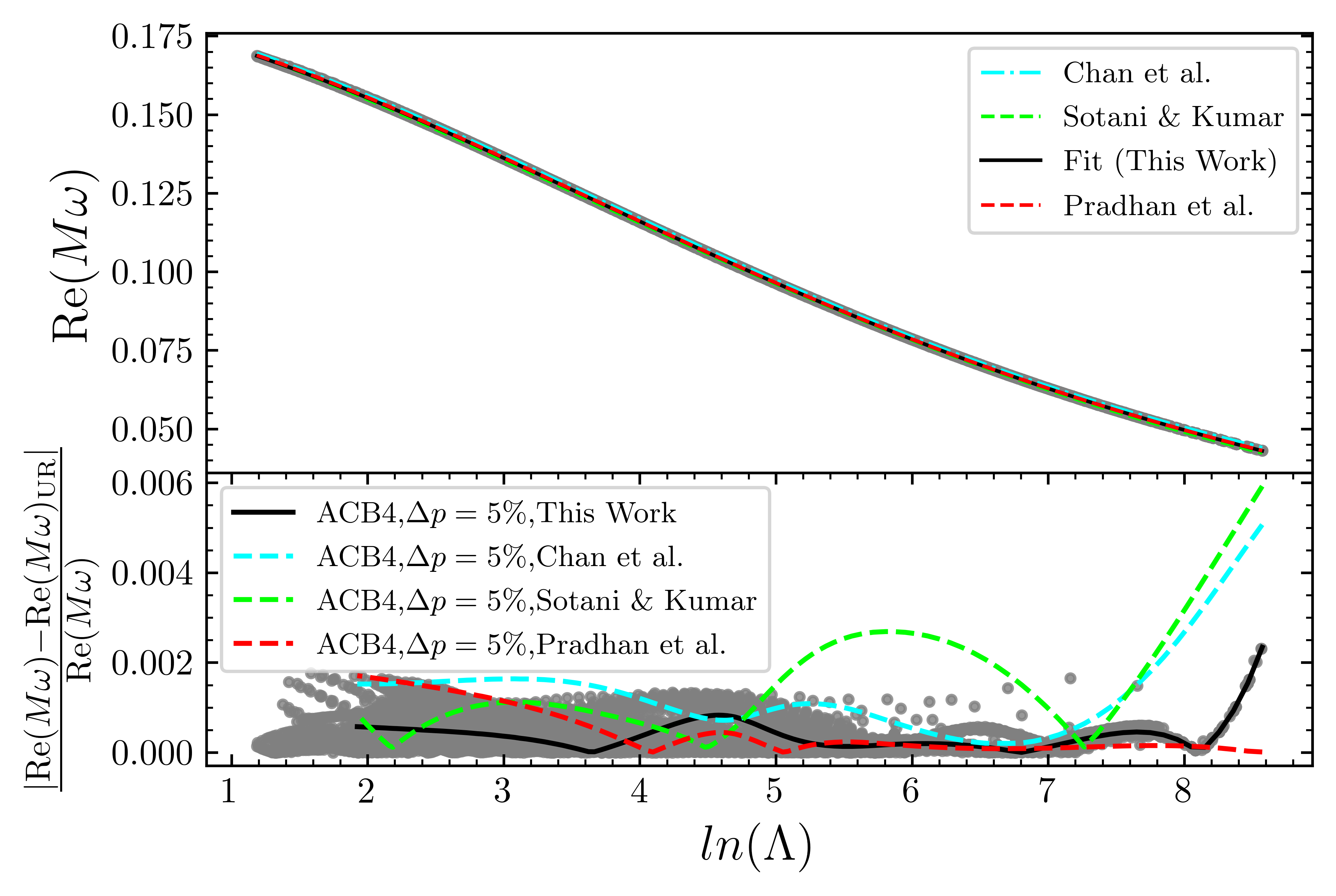}
    \caption{ (Upper panel)$f-\mathrm{Love}$ relation resulting from consideration of a wide range of hybrid EOSs along with the different $f-\mathrm{Love}$ relations from the literature includes URs from Chan et al.~\citep{Chan2014}, Sotani \& Kumar~\citep{Sotani2021}, and ~Pradhan et al.,~\citep{Pradhan2023b}. (lower panel) The relative error on the $\mathrm{Re} (M\omega)$ for the corresponding fit relation has also been displayed. Additionally,  different URs and the resulting error due to different URs on the $M\omega$ of a representative hybrid EOS ACB4,$\Delta p=5\%$, are also shown. }
    \label{fig:f_love_new}
\end{figure}

Additionally, there are also URs, including the mass-scaled  $f$-mode parameters and tidal deformability parameters (referred to as $f$-Love relation), which in a binary system include the necessary correction due to stellar mode excitation~\citep{Pradhan2023,Chan2014} or inversely constrain $f$-mode parameters from $M$ and $\Lambda$ measurements~\citep{Wen}. We have displayed the universal relation among Re($M\omega$) and $\Lambda$ in ~\cref{fig:f_love_new}. In addition to the URs obtained in this work, we have also displayed the $f$-Love relation from other relevant works ~\citep{Chan2014,Sotani2021,Pradhan2023b}.  The $f$-Love relations $M\omega_R-\Lambda$ and $M/\tau-\Lambda$ are given in ~\cref{eqn:remomega_lambda} and ~\cref{eqn:immomega_lambda} respectively. We tabulate the fit parameters for the relations~\cref{eqn:remomega_lambda} and~\cref{eqn:immomega_lambda} in~\cref{tab:mmomeg_lambda}.  URs given in~\cref{eqn:remomega_lambda,eqn:immomega_lambda} can be used to simultaneously infer the $M$ and  $\Lambda$ from the detection of ($f,\tau$),  and then using the $C-\Lambda$ relation~\citep{Pradhan2023} $R$ can be reconstructed. 

\begin{equation}\label{eqn:remomega_lambda}
    M\omega_R=\sum_{j=0}^{5} c_{R,j} z^j
\end{equation}
\begin{equation}\label{eqn:immomega_lambda}
   M\omega_I=\frac{M}{\tau}=\sum_{j=0}^{5} c_{I,j} z^j
\end{equation}
where $z=\log(\Lambda)$.
\\

\begin{table}
    \centering
    
    \begin{tabular}{|p{0.1\linewidth}|p{0.25\linewidth}|p{0.1\linewidth}|p{0.25\linewidth}|}
    \hline
 \multicolumn{2}{|c|}{$M\omega_R-\log{(\Lambda)}$}& \multicolumn{2}{|c|}{$M\omega_I-\log{(\Lambda)}$} \\
 \hline
         $c_{R,0}$&$0.182\pm 5.48\times 10^{-5} $ &
         $c_{I,0}$&(3.437$\pm 0.005) \times 10^{-5}$ \\
         \hline
         $c_{R,1}$&-$(6.110\pm 0.006)\times10^{-3}$  &
         $c_{I,1}$& (3.921$\pm 0.004) \times 10^{-5}$\\
         \hline
         $c_{R,2}$&-$(4.594\pm 0.03)\times10^{-3}$  &
         $c_{I,2}$& (-9.988$\pm 0.009) \times 10^{-6}$ \\
         \hline
         $c_{R,3}$&$(6.066\pm 0.009)\times10^{-4}$   &
         $c_{I,3}$& $(1.007\pm 0.032)\times10^{-7}$  \\
         \hline
         $c_{R,4}$&-$(2.614\pm 0.006)\times10^{-5}$&
         $c_{I,4}$& $(1.235\pm 0.008)\times10^{-7}$ \\
         \hline
         $c_{R,5}$&$(2.228\pm 0.03)\times10^{-7}$  &
         $c_{I,5}$& -$(7.754\pm 0.01)\times10^{-9}$ \\
         \hline
    \end{tabular}
    \caption{Fit parameters for $M\omega$ with $\log{(\Lambda)}$ as fit parameters given in ~\cref{eqn:remomega_lambda,eqn:immomega_lambda}.}
    \label{tab:mmomeg_lambda}
\end{table}

\begin{figure}

  \includegraphics[width=\linewidth]{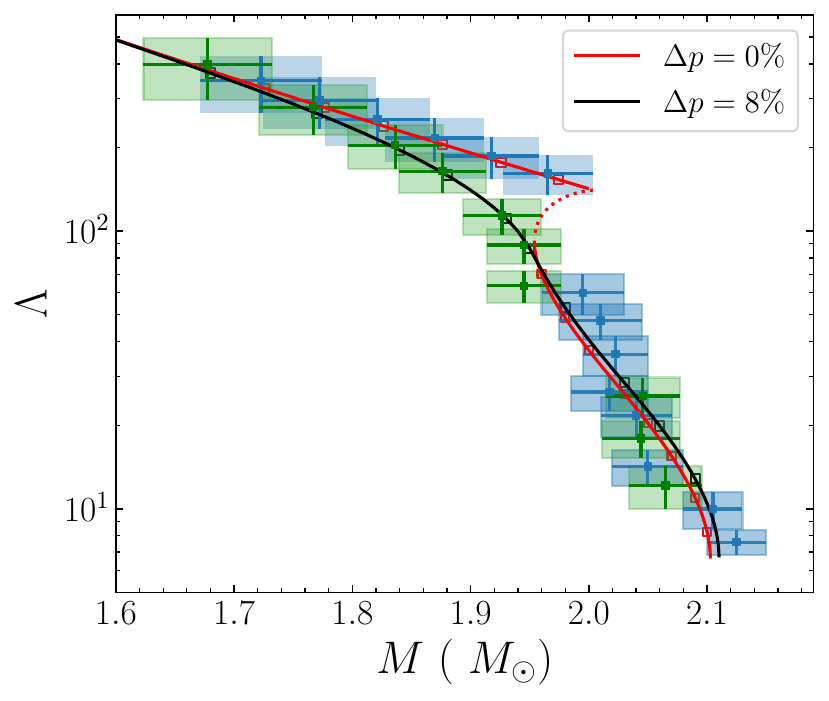}
     
   \caption{Recovered $M-\Lambda$ using URs 
   ~\cref{eqn:remomega_lambda,eqn:immomega_lambda} for a few randomly chosen configurations with the assumption of precise measurement of $f$-mode parameters. The injections are shown with empty squares (in red for $\Delta p=0\%$ and black for $\Delta p=8\%$). The uncertainties are shown in  blue (green) for EOS with $\Delta p=0\%$ ($\Delta p=8\%$).}
   \label{fig:mlambda_recovery_ur}
    \end{figure}

In general, the URs involving $\Lambda$ are of great use and also have fewer uncertainties.  We  reconstruct the $M$ and $\Lambda$ using the URs in~\cref{eqn:remomega_lambda,eqn:immomega_lambda} for a few randomly chosen configurations and display the errors in the $M-\Lambda$  resulting  from the uncertainties in URs  in~\cref{fig:mlambda_recovery_ur}. From ~\cref{fig:mlambda_recovery_ur}, one can conclude that the URs recover the stellar properties well within certain uncertainties. We notice that within the uncertainty in ~\cref{fig:mlambda_recovery_ur}, one may find difficulty  distinguishing  the value of $\Delta p$ as there are overlap regions for the recovered $M-\Lambda$ corresponding to $\Delta p=8\%$ and $\Delta p=0\%$. Additionally, we also notice that the recovered $M-\Lambda$ for the twin companions are distinguishable.

\section{Hybrid EOS models used in this work and Universal Relations}\label{app:app_B}

\begin{figure}
    \centering
    \includegraphics[width=\linewidth]{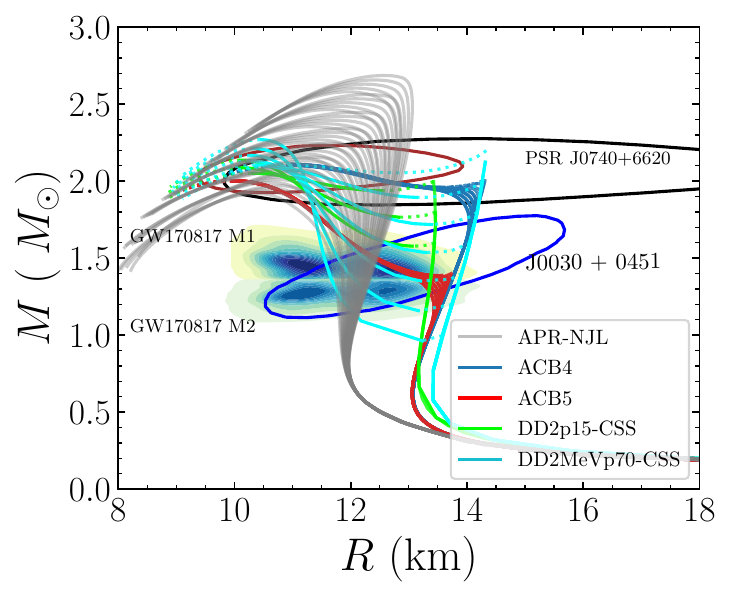}%
    \caption{ Similar to ~\cref{subfig:mr}, but showing  $M-R$ relations for all the hybrid EOSs included in obtaining the URs.}
    \label{fig:MR_new}
\end{figure}

In the main text, we present the equation of state (EOS) and the mass-radius ($M-R$) relations for the ACB4 EOS model, which is central to our discussions. In \cref{fig:MR_new}, we showcase all the hybrid EOS models considered for obtaining the universal relations (URs). These include: (i) APR-NJL model from ~\citep{Ayriyan2021b}, (ii) Hybrid EOS models based on the DD2 hadronic EOS and constant speed of sound quark matter model  from ~\citep{David2021}, labeled as DD2p15-CSS and DD2MeVp70-CSS in ~\cref{fig:MR_new}, (iii) The ACB4 and ACB5  hybrid models from~\citep{Paschalidis2018}.

\begin{figure*}
\centering

\subfloat[]{%
  \includegraphics[width=0.44\textwidth]{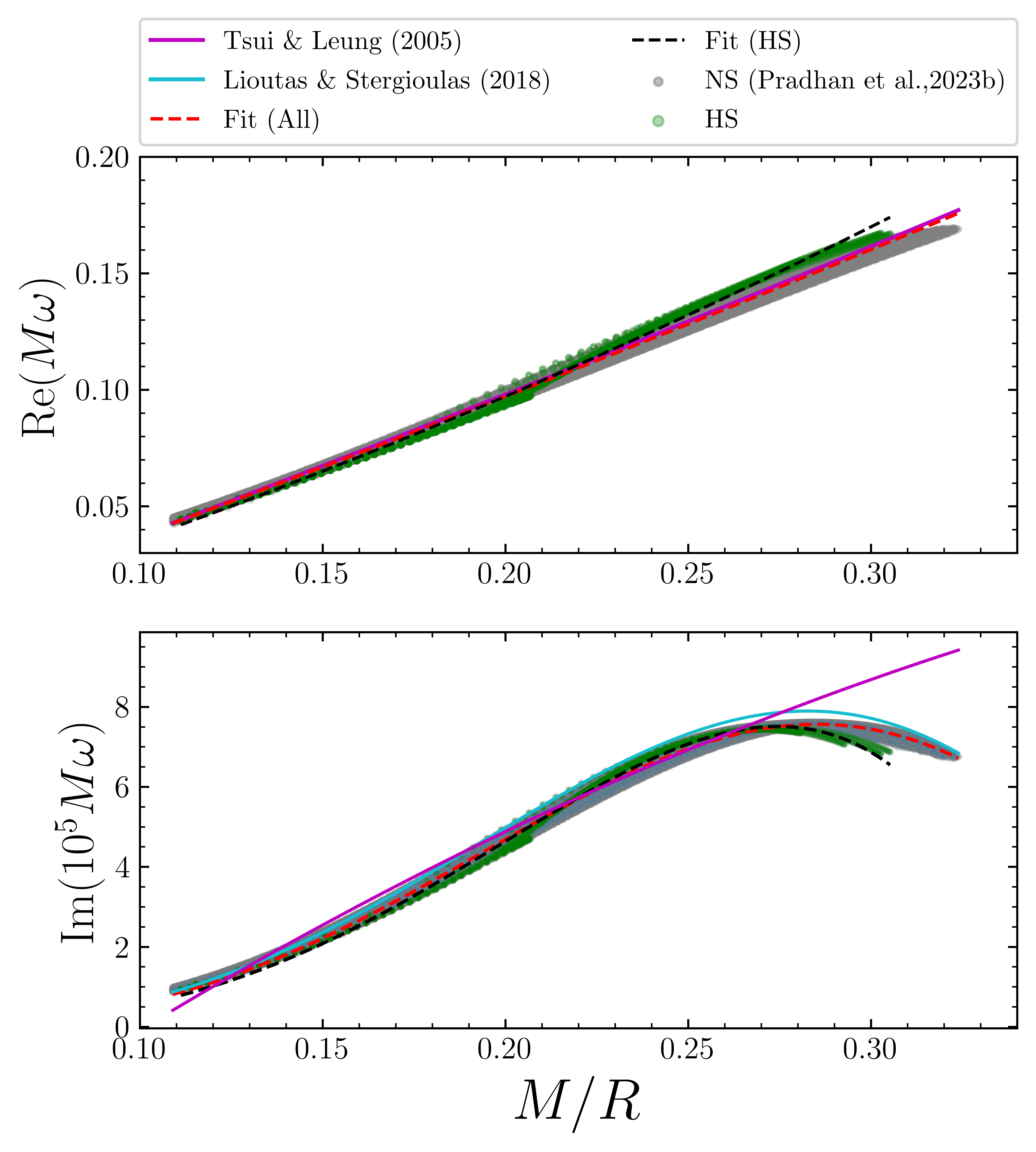}%
  \label{subfig:UR_compare_v1}%
}
\subfloat[]{%
  \includegraphics[width=0.44\textwidth]{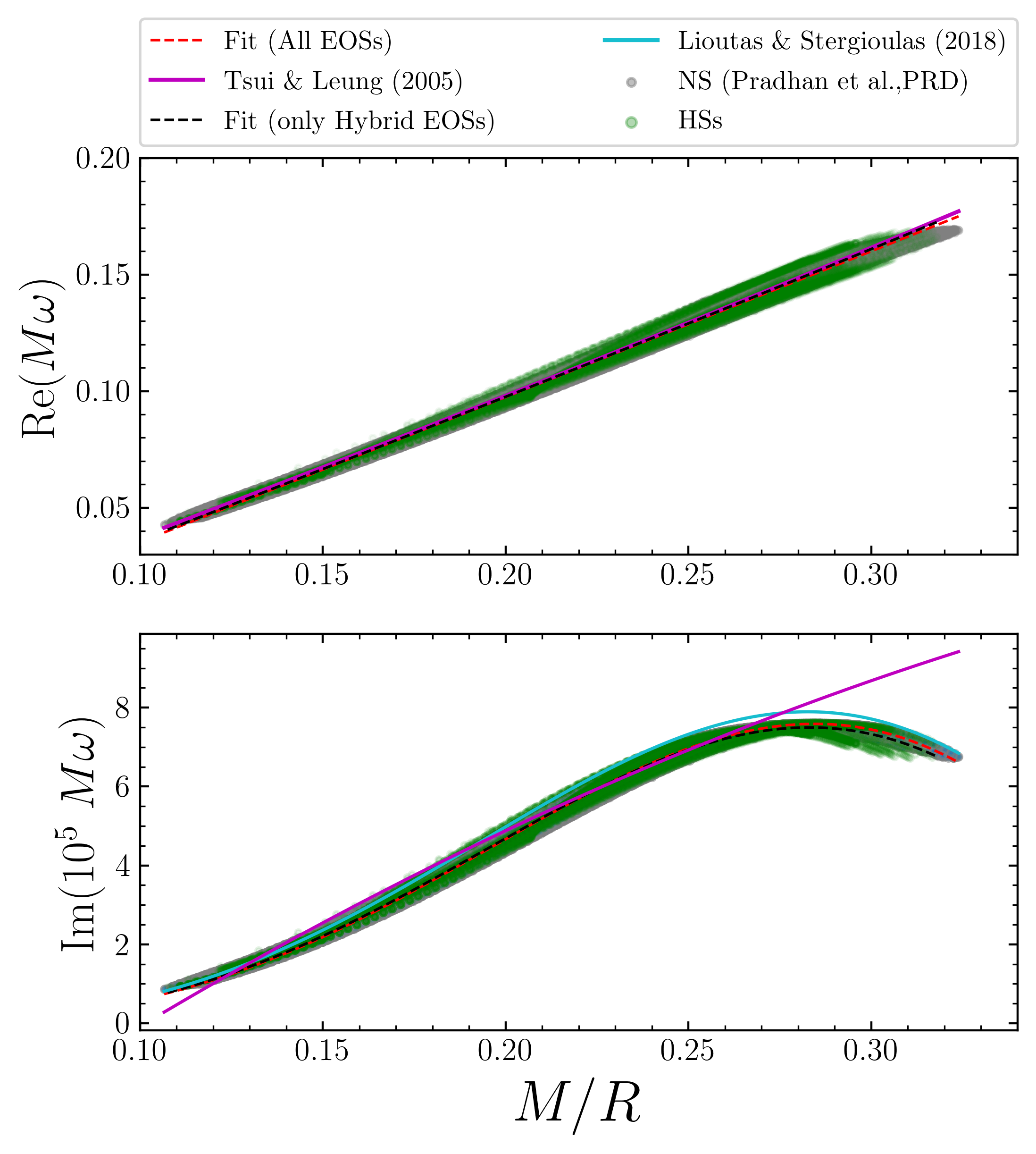}%
  \label{subfig:UR_compare_new}%
}
\caption{(a) Universal relation among $M\omega$ and stellar compactness $M/R$, along with the fit relations. For comparison, fit relations obtained from considering only the ACB4 and ACB5 hybrid EOSs (labeled as Fit (HS) ), all EOS models (hybrid EOSs and nucleonic, polytropic, hyperonic; labelled as Fit (All) ) along with URs from different works have also been shown. (b) Same as (a), but the a large family of hybrid EOSs are considered for analysing the URs (see, ~\cref{fig:MR_new} for the $M-R$ relations). }
\end{figure*}

Considering only the ACB4 and ACB5  hybrid EOSs (see, ~\cref{fig:EOS_MR,fig:EOS_MR_ACB5}), the URs are shown in~\cref{subfig:UR_compare_v1}. In ~\cref{subfig:UR_compare_v1} , we present two distinct fit relations: ``Fit (All)" derived by considering all the EOSs from ~\citep{Pradhan2023b} along with the ACB paremetrized hybrid EOSs, and ``Fit (HS) " obtained using only the ACB4 and ACB5 hybrid EOSs. Though consideration of only HSs results in a different fit relation in the higher compactness region, the theoretical values are within the resulting uncertainty band of NSs. Moreover, considering a wide range of hybrid EOS models, as depicted in \cref{fig:MR_new}, leads to a theoretical uncertainty of the scaled mode characteristics covering a broad span (see ~\cref{subfig:UR_compare_new}), including the uncertainty band of the NS models. The inclusion of a large family of hybrid EOSs results in a comparatively larger uncertainty at higher compactness regions. This broader uncertainty range near the higher compactness region affects the $M-R$ recovery, introducing a larger uncertainty in the recovered $M-R$ in ~\cref{sec:asteroseismology} for massive HSs. However, the uncertainties associated with the URs involving the tidal deformability remain minimal.

\bsp	
\label{lastpage}
\end{document}